\documentclass[12pt,a4paper]{article}
\usepackage[utf8]{inputenc}
\usepackage[left=18mm,top=20mm,right=18mm,bottom=15mm]{geometry}
\usepackage[english]{babel}
\usepackage{mathrsfs}
\usepackage{amsmath}
\usepackage{amsfonts}
\usepackage{amssymb}
\usepackage{makeidx}
\usepackage{graphicx}
\usepackage{physics}
\usepackage{float} %Include figure filesusepackage{graphicx} %Include figure files
\usepackage{subfloat}
\usepackage{multicol}
\usepackage{subcaption}
\usepackage{color}
\usepackage[dvipsnames]{xcolor}
\usepackage{enumerate}
\usepackage[colorlinks]{hyperref}
\hypersetup{
  colorlinks,
  citecolor=red,
  linkcolor=red,
  urlcolor=violet}
\pagecolor[rgb]{1.0,0.99,1.0}
\usepackage{wrapfig}   
\usepackage{paralist} 
\usepackage{soul}

\usepackage{authblk}

\title{More than one Author with different Affiliations}
\author[1]{Jitendra Pal\thanks{jpal1@ph.iitr.ac.in}}
\author[1]{Hemant Rathi\thanks{hrathi@ph.iitr.ac.in}}
\author[2,3]{Arindam Lala\thanks{arindam.physics1@gmail.com}}
\author[1]{Dibakar Roychowdhury\thanks{dibakar.roychowdhury@ph.iitr.ac.in}}
\affil{Department of Physics, Indian Institute of Technology Roorkee, Roorkee
247667 Uttarakhand, India}
\affil[2]{Department of Physics, Indian Institute of Technology Madras, Chennai
600036 Tamil Nadu, India}
\affil[3]{Institute of Physics, sachivalaya Marg, Bhubaneswar, Odisha 751005, India \\
\textbf{\&}
\\
Homi Bhabha National Institute,Training School Complex, Anushakti Nagar, Mumbai 400085, India}

\begin{document}
\date{}
%%%%%%%%%%%%%%%%%%%%
\title{{\bf{\Large Non-chaotic dynamics for Yang-Baxter deformed $\text{AdS}_{4}\times
\text{CP}^{3}$ superstrings}}}
%%%%%%%%%%%%%%%%%%%%
\iffalse
\author{
 {\bf {\normalsize Dibakar Roychowdhury}$
$\thanks{E-mail:  dibakarphys@gmail.com, dibakarfph@iitr.ac.in}}\\
 {\normalsize  Department of Physics, Indian Institute of Technology Roorkee,}\\
  {\normalsize Roorkee 247667 Uttarakhand, India}
\\[0.3cm]
}
\fi
%\date{}
%%%%%%%%%%%%%%%%%%%%
\maketitle
%%%%%%%%%%%%%%%%%%%%
\begin{abstract}\label{abstract}
We explore a novel class of Yang-Baxter deformed AdS$_{4}$ $\times$ CP$^{3}$
backgrounds [Jour. High Ener. Phys. \textbf{01} (2021) 056] which exhibit a non-chaotic dynamics for (super)strings propagating over it. We explicitly use the \textit{Kovacic's algorithm} in order to establish non-chaotic dynamics of string $ \sigma $ models over these deformed backgrounds. This analysis is complemented with numerical techniques whereby we probe the classical phase space of these (semi)classical strings and calculate various chaos indicators, such as, the Poincar\'{e} sections and the Lyapunov exponents. We find compatibility between the two approaches. Nevertheless, our analysis does not ensure integrability; rather, it excludes the possibility of non-integrability for the given string embeddings.
\end{abstract}
%%%%%%%%%%%%%%%%%%%%%%%%%%%%%%%%%%%%%%%%
\newpage
%%%%%%%%%%
\tableofcontents
%%%%%%%%%%%%%%%%%%%%
\section{Introduction and summary}\label{intro}

Understanding the chaotic behaviour \cite{Zayas:2010}-\cite{Banerjee:2018ifm} and the associated 
non-integrable structure in various examples of gauge/gravity correspondence 
\cite{Maldacena:1997re}-\cite{Witten:1998qj} has been an outstanding problem for past couple of 
decades. While in most of these cases one encounters a chaotic motion, there have been some
handful of examples that confirm non-chaotic behaviour of the embedded super strings and hence rules out the possibility of non-integrable dynamics in the stringy phase space.

\vspace{5pt}

Non-chaotic dynamics are therefore always special in holographic dualities. The central idea
behind these analyses is to probe the classical phase space configuration of (semi-)classical
strings with various chaos indicators. These indicators ensure whether the phase space allows
a Kolmogorov–Arnold–Moser (KAM) tori and thereby (quasi-)periodic orbits 
\cite{Zayas:2010}-\cite{Basu:2011b}. Identification of these orbits in the first place, is the
key step towards unveiling an integrable structure associated with the classical phase space.

\vspace{5pt}

On the other hand, one can use the notion of Kovacic's algorithm to analytically check
the Liouvillian (non-)integrability criteria for a classical $2d$ sigma model over general
backgrounds based on a set of \textit{necessary but non sufficient} rules
\cite{Kovacic:1986}-\cite{Kovacic:2005}. In this paper, we use both these methods to explore
classical  (non-)chaotic dynamics of the associated sigma models in the stringy phase space.\footnote{It must be emphasized that, parallel to the Kovacic's algorithm, there exist other approaches to check the (non-)integrability of the sigma models, such as the S-matrix factorization \cite{Wulff:2017lxh,Wulff:2017vhv,Wulff:2019tzh,Giataganas:2019xdj}.}

\vspace{5pt}

Following the holographic duality \cite{Maldacena:1997re}-\cite{Witten:1998qj}, one can argue
that these semi-classical strings are dual to a class of single trace operators in the large $N$ limit
of the dual QFT. This would therefore enable us to conjecture about the integrability of the dual
QFT at strong coupling. It must be stressed that, examples of integrable superstring sigma
models within the holographic dualities are scarce. In fact, the absence of any systematic
procedure to construct Lax pairs for these two-dimensional field theories makes our tasks even 
more challenging. However, so far there are some handful of examples starting with AdS$_{5}
\times S^{5}$ and AdS$_{4}\times $ CP$^{3}$ where the classical integrability can be established
by means of Lax pair \cite{Bena:2003wd}-\cite{Zarembo:2010sg}.  On the other hand, it is equally
interesting to look for integrable models which are deformations of the original sigma models. Along
this line, $\beta$-deformations (a marginal deformation) of the $\mathcal{N}=4$ super-Yang-Mills
(SYM) theory \cite{Lunin:2005jy}, which is dual to the type IIB super-string theory on AdS$_{5}
\times S^{5}$, was studied in \cite{Frolov:2005ty}-\cite{Frolov:2005dj}. The deformed model was
found to be integrable \cite{Frolov:2005dj} for real deformation parameters and non-integrable \cite{Giataganas:2013dha} for complex deformation parameters.

\vspace{5pt}

The purpose of the present paper is to apply these concepts to a novel class of Yang-Baxter (YB)
deformed \cite{Klimcik:2002zj}-\cite{Matsumoto:2015uja} backgrounds those were obtained
until  recently by the authors in \cite{Negron:2018btz}-\cite{Rado:2021mji}. These are the
deformations of the original AdS$_{4}$ $\times$ CP$^{3} $ background \cite{Aharony:2008ug}
where the deformation is generated through classical $ r $-matrices satisfying the YB equation. However, unlike the undeformed case
\cite{Arutyunov:2008if}-\cite{Sorokin:2010wn}, the integrable structures associated with these
deformed class of backgrounds are yet to be confirmed through systematic analyses. 

\vspace{5pt}

\vspace{5pt}

Classical $ r $-matrices satisfying modified classical Yang-Baxter equation (mCYBE) \cite{Matsumoto:2015jja,Matsumoto:2014cja} have been applied to symmetric cosets
\cite{Delduc:2013fga} as well as AdS$_5$ $\times$ $S^5 $ super-cosets 
\cite{Delduc:2013qra}-\cite{Arutyunov:2013ega}. For the later case, the type IIB equations were 
confirmed until recently \cite{Hoare:2018ngg}. On the other hand, Abelian $ r $-matrices
satisfying CYBE were applied to AdS$_5$ $\times$ $S^5 $ sigma models in
\cite{Kawaguchi:2014qwa} which were further generalized for the non-Abelian case in 
\cite{Orlando:2016qqu}. For classical $ r $-matrices satisfying CYBE, the resulting background is
found to satisfy type IIB supergravity equations of motion.
\iffalse
\footnote{In order to satisfy type IIB
equations, the $ r $-operator has to be \emph{unimodular} \cite{Borsato:2016ose}. On the other
hand, for non-unimodular $ r $-operators, one solves modified supergravity equations 
\cite{Arutyunov:2015qva}-\cite{Arutyunov:2015mqj}. In case of classical YB deformations, the
non-Abelian bosonic jordanian $ r $-operators are non-unimodular. These are therefore associated
with modified supergravity solutions \cite{Hoare:2016hwh}. Abelian $ r $-operators, on the other
hand, are always unimodular which are threfore associated with usual type II supergravity
backgrounds \cite{vanTongeren:2015soa}, \cite{Matsumoto:2014gwa}-\cite{Matsumoto:2014cja}.} 
\cite{Matsumoto:2014nra}-\cite{Matsumoto:2015uja}.
\fi

\vspace{5pt}

Motivated by these AdS$_5$ $\times$ $S^5 $ examples, abelian $ r $-matrices satisfying CYBE
have been applied to AdS$_4$ $\times$ CP$^3 $ sigma models until very recently 
\cite{Negron:2018btz}-\cite{Rado:2021mji}. In their construction, the authors consider various
YB deformations of the AdS$_{4} $ subspaces and/or the internal CP$^{3} $ manifold. These
result into a class of deformed ABJM models (as dual descriptions) which we summarise below.

\vspace{5pt}

Depending on the type of YB deformations, one eventually generates a class of gravity duals 
\cite{Negron:2018btz}-\cite{Rado:2021mji} for (1) $\beta$-deformed ABJM, (2) Noncommutative
ABJM, (3) Dipole deformed ABJM and (4) Nonrelativistic ABJM. It is worth mentioning that three
parameter $ \beta $-deformed backgrounds can also be obtained following a TsT
(T-duality--shift--T-duality) transformation of AdS$_4$ $\times$ CP$^3 $ \cite{Imeroni:2008cr}.
On a similar note, a three parameter dipole deformation as well as gravity duals for noncommutative
ABJM were also obtained by applying TsT transformations on AdS$_4$ $\times$ CP$^3 $ backgrounds
\cite{Imeroni:2008cr}. Moreover, the TsT transformation on the AdS$_4$ $\times$ CP$^3 $
background generating the gravity dual of the nonrelativistic ABJM has also been found in
\cite{Rado:2021mji}. These guarantee that all these YB deformed backgrounds are string backgrounds
in the type IIA supergravity.

\vspace{5pt}

In the present paper, we consider (semi)classical string dynamics for each of these deformed
backgrounds and calculate their respective chaos indicators, namely, the Poincar\'{e} section and
the Lyapunov exponent ($\lambda$) \cite{Zayas:2010}-\cite{Basu:2011b}. For an integrable
dynamical system that does not show chaos, the $2N$ dimensional phase space consists of $N$ 
dimensional hypersurfaces known as KAM tori. In these dynamical systems the equations of motion 
describe a flow in the phase space which are indeed nicely foliated trajectories. However, in order
to make the analysis simpler, a lower dimensional slicing of the KAM tori is chosen. This later
hypersurface is known as the Poincar\'{e} section. The flow trajectories then continuously cross the 
Poincar\'{e} section. When chaos sets in, the nice shape of the KAM tori is destroyed. On the other
hand, the Lyapunov exponent ($\lambda$) is an essential tool to determine the chaotic behaviour
of a dynamical system. It is the rate of the exponential separation of initially close trajectories in the
phase space of the system with time. When the system is non-chaotic, $\lambda$ decays to zero
with time. Whereas for a chaotic system, the initial separation between two nearby trajectories
grows exponentially. A non-zero positive value of $\lambda$ is usually an indication of chaos.

\vspace{5pt}

In our analyses, we find no indications of chaotic dynamics of the strings; the shapes of the KAM tori
are never distorted and the Lyapunov exponents decay to zero over time. By implementing numerical algorithm, we test these latter results for various possible values of the string energy as well. Our numerical analyses
are substantiated by analytical computations. The analytical calculations make use of the
Kovacic's algorithm which determines the Liouvillian (non-)integrability of a homogeneous linear
second order ordinary differential equation with polynomial coefficients 
\cite{Kovacic:1986,Saunders:1981,Kovacic:2005,Nunez:2018qcj}. It must be stressed that, in our analysis, the Kovacic's algorithm rules out the possibility of non-integrability. The system is likely to be integrable. However, in our case to ensure the integrability we need to find the appropriate Lax pair which is not the focus of the present article. More details about the
analytical and numerical methodologies are provided in Appendices \ref{Kova} and \ref{app:num},
respectively.

\vspace{5pt}

The organization for the rest of the paper is as follows. In Section \ref{sec:set}, we present the
preliminary requisites to perform our analyses for the rest of the paper. In Section \ref{Results},
we apply the analytical as well as numerical algorithms to look for indications of chaotic behaviours
of the string sigma models for each of the four examples listed above. Finally, we conclude in
Section \ref{Conclusion}. The two Appendices \ref{Kova} and \ref{app:num} describe the
analytical and numerical methods that have been used in our analyses. The additional two
Appendices \ref{cofT:BD} and \ref{der:NR} collect several mathematical expressions that
appear in the main text of the article.

%%%%%%%%%%%%%%%%%%%%%%%%%%%%%%%%
\section{Basic set up}\label{sec:set}
The starting point of our analysis will be the classical $2d$ string sigma model which, in the conformal
gauge, can be written as \cite{Polchinski:1998rq}
%%%%
\begin{equation}
\label{POL}
S_{P}=-\frac{1}{2}\int \dd\tau\dd\sigma \left( \eta^{ab}G_{MN}+\epsilon^{ab}
B_{MN}\right)\partial_{a}X^{M}\partial_{b}X^{N} \; ,
\end{equation}
%%%%
where $\displaystyle \eta_{ab}=\text{diag}\left(-1,1\right)$ is the world-sheet metric with
world-sheet coordinates $(\tau,\sigma)$. We choose the following convention for the Levi-Civita
symbol: $\epsilon^{\tau\sigma}=-1$. Note that, the above action (\ref{POL}) is the Polyakov
action in the presence of non-trivial $B$-field.

The conjugate momenta corresponding to the target space coordinates $X^{\mu}$ can be
computed from the action (\ref{POL}) as
%%%%
\begin{equation}\label{con:mom}
p_{\mu}=\frac{\partial \mathcal{L}_{P}}{\partial\dot{X}^{\mu}}=
G_{\mu\nu}\partial_{\tau}X^{\nu}+B_{\mu\nu}\partial_{\sigma}X^{\nu} \, .
\end{equation}
%%%%

The Hamiltonian of the system can be written as
%%%%
\begin{equation}\label{Hamil}
\mathcal{H} = p_{\mu}\partial_{\tau}X^{\mu}-\mathcal{L}_{P}=
\frac{1}{2}G_{\mu\nu}\qty(\partial_{\tau}X^{\mu}\partial_{\tau}X^{\nu}
+\partial_{\sigma}X^{\mu}\partial_{\sigma}X^{\nu}) \, .
\end{equation}
%%%%
Note that, the Hamiltonian (\ref{Hamil}) is indeed equal to the $(\tau,\tau)$ component
$T_{\tau\tau}$ of the energy-momentum tensor $T_{ab}$ whose general expression can
be derived from the action (\ref{POL}) as
%%%%
\begin{equation}\label{EM:gen}
T_{ab} = \frac{1}{2}\qty(G_{\mu\nu}\partial_{a}X^{\mu}\partial_{b}X^{\nu}
-\frac{1}{2}h_{ab}h^{cd}G_{\mu\nu}\partial_{c}X^{\mu}\partial_{d}X^{\nu}) \, ,
\end{equation}
%%%%
where $h_{ab}=e^{2\omega\qty(\tau,\sigma)}\eta_{ab}$ in the conformal gauge
\cite{Ishii:2021asw}.

The Virasoro constraints imply that
%%%%
\begin{align}\label{Vir:EM}
\begin{split}
T_{\tau\tau} = T_{\sigma\sigma} &=0 \, ,  \\
T_{\tau\sigma} = T_{\sigma\tau} &=0 \, .
\end{split}
\end{align}
%%%%

%%%%%%%%%%%%%%%%%%%%%%%%%%%%%%%%%
\section{Main results: Analytical and numerical}\label{Results}
The purpose of this section is to elaborate on the key analytical as well as numerical steps to
check (non-)chaotic dynamics of the string $\sigma$-models within Yang-Baxter (YB) deformed ABJM
theories those are in accordance to the algorithms described in Appendices \ref{Kova} and
\ref{app:num}, respectively. Below, we describe them in detail taking individual examples of
the YB deformed ABJM model.
%%%%%%%%%%%%%%%%%%%%%%%%%%%%%%%%%%
\subsection{$\beta$-deformed ABJM}\label{ABJM:BD}
The Yang-Baxter (YB) deformed background dual to $\beta$-deformed ABJM is obtained by
deforming the CP$^3$ subspace using Abelian $r$-matrices\footnote{The form of the
$r$-matrix that leads to the three-parameter deformed background (\ref{met:BD}) is chosen
as \cite{Negron:2018btz}
\begin{equation*}
r =~ \hat{\gamma}_{1} \mathbf{L} \wedge \mathbf{M}_{3}
+ \hat{\gamma}_{2} \mathbf{L}_{3} \wedge \mathbf{M}_{3}
+ \hat{\gamma}_{3} \mathbf{L}_{3} \wedge \mathbf{L} \, ,
\end{equation*}
where $\mathbf{L}=-1/\sqrt{3}\,\mathbf{L}_{8}+\sqrt{2/3}\,\mathbf{L}_{15}$ and
$\mathbf{L}_{3} \, , \mathbf{L}_{8} \, , \mathbf{L}_{15} \, , \mathbf{M}_{3}$ $\in$
$\mathfrak{su}(4)\oplus \mathfrak{su}(2)$ are Cartan generators.}\cite{Negron:2018btz}
which results in the following space-time line element
%%%%
\begin{align}\label{met:BD}
\begin{split}
\dd s_{R \times CP^3}^{2} &=-\frac{1}{4}\dd t^{2}+\dd \xi^{2}+\frac{1}{4}
\cos ^{2} \xi\left(\dd \theta_{1}^{2}+\mathcal{M} \sin ^{2} \theta_{1} 
\dd\varphi_{1}^{2}\right)+\frac{1}{4} \sin ^{2} \xi\left(\dd \theta_{2}^{2}+
\mathcal{M} \sin^{2} \theta_{2} \dd \varphi_{2}^{2}\right)    \\
&+\mathcal{M} \cos^{2} \xi \sin^{2} \xi\left(\dd\psi+\frac{1}{2}\cos\theta_{1}
\dd\varphi_{1}-\frac{1}{2} \cos\theta_{2} \dd \varphi_{2}\right)^{2}   \\
&+\mathcal{M} \sin^{4} \xi \cos^{4} \xi \sin^{2} \theta_{1} \sin^{2} 
\theta_{2}\left(\hat{\gamma}_{1} \dd \varphi_{1}+\hat{\gamma}_{2} 
\dd\varphi_{2}+\hat{\gamma}_{3} d \psi\right)^{2}  \, .
\end{split}
\end{align}
%%%%

Notice that, in writing the metric (\ref{met:BD}) we switch off the remaining coordinates of
the $AdS_{4}$. Here $\hat{\gamma}_{i}$ ($i=1,2,3$) are the YB deformation parameters.

The corresponding NS-NS 2-form field is given by
%%%%
\begin{align}\label{ns:BD}
\begin{split}
B &=-\mathcal{M} \sin ^{2} \xi \cos ^{2} \xi\bigg[\frac{1}{2}(2 \hat{\gamma}_{2} 
+\hat{\gamma}_{3} \cos \theta_{2}) \cos ^{2} \xi \sin ^{2} \theta_{1} \dd \psi
\wedge \dd \varphi_{1}     \\
&+\frac{1}{2}\left(-2 \hat{\gamma}_{1}+\hat{\gamma}_{3} \cos \theta_{1}\right)
\sin ^{2} \xi \sin ^{2} \theta_{2} d \psi \wedge d \varphi_{2}+\frac{1}{4}
\bigg(\hat{\gamma}_{3} \sin ^{2} \theta_{1} \sin ^{2} \theta_{2}     \\
&+(2 \hat{\gamma}_{2}+\hat{\gamma}_{3} \cos \theta_{2}) \cos ^{2} \xi \sin ^{2}
\theta_{1}\cos\theta_{2}+(-2 \hat{\gamma}_{1}+\hat{\gamma}_{3}\cos\theta_{1})
\sin^{2}\xi\sin^{2}\theta_{2}\cos\theta_{1}\bigg) \dd \varphi_{1}
\wedge\dd\varphi_{2}\bigg]  \, ,
\end{split}
\end{align}
%%%%
where
%%%%
\begin{align}\label{funDef:BD}
\begin{split}
\mathcal{M}^{-1} &=\hspace{1mm}1+ \sin^{2}\xi\cos^{2}\xi
\bigg(\hat{\gamma}_3^2\sin^{2} \theta_{1}\sin^{2}\theta_{2}
+(2 \hat{\gamma}_2+\hat{\gamma}_3\cos{\theta_2})^2\cos^{2}\xi
\sin^{2}\theta_{1}    \\
&+ (-2\hat{\gamma}_1+\hat{\gamma}_3\cos{\theta_1})^2\sin^2{\xi}
\sin^2{\theta_2} \bigg)  \, .
\end{split}
\end{align}
%%%%

Next we consider the winding string ansatz given by
%%%%
\begin{align}\label{ansatz:BD}
t &=t(\tau ) \, , & \theta_1 &= \theta_1(\tau) \, , & \theta_2 &=\theta_2(\tau) \, ,
&   \xi &=\xi(\tau) \, ,   \\[5pt]\nonumber
\phi_1 &= \alpha_2\sigma \, ,  & \phi_2 &=
\alpha_4\sigma \, ,  & \psi &=\alpha_6\sigma \, ,
\end{align}
%%%%
where $\alpha_{i}$ ($i=2,4,6$) are the winding numbers.

%In our following analyses \textendash{} analytical as well as numerical \textendash{} we make
%the winding numbers $\alpha_{i}=1$ ($i=2,4,6$).

Using the above ansatz (\ref{ansatz:BD}), the Lagrangian density in the Polyakov action
(\ref{POL}) can be written as
%%%%
\begin{subequations}
\begin{alignat}{2}
\begin{split}
L_{P} &=~ -\frac{1}{2} \Bigg[\frac{1}{4} \dot{t}^{2}-\dot{\xi}^{2}
			-\frac{1}{4}\qty(\dot{\theta}_{1}^{2} \cos^{2}\xi +\dot{\theta}_{2}^{2}
			\sin^{2}\xi ) +\frac{\mathcal{M}	\phi_1'^2}{4}\cos^{2}\xi
			\Big( \sin^{2}\theta_{1}+\sin^{2}\xi \cos^{2}\theta_{1}   \\
			& +4\hat{\gamma}_{1}^{2}\sin^{2}\theta_{1}\sin^{2}\theta_{2}
			\sin^{4}\xi\cos^{2}\xi \Big)+\frac{\mathcal{M}	\phi_2'^2}{4}\sin^{2}\xi
			\Big( \sin^{2}\theta_{2}+\cos^{2}\xi\cos^{2}\theta_{2}+4\hat{\gamma}_{2}^{2} 
			\sin^{2}\theta_{1}    \\
			& \times \sin^{2}\theta_{2}\sin^{2}\xi\cos^{4}\xi \Big) +\mathcal{M}	\psi'^2
			\sin^{2}\xi \cos^{2}\xi \Big( 1+\hat{\gamma}_{3}^{2} \sin^{2}\theta_{1}\sin^{2}
			\theta_{2}\sin^{2}\xi\cos^{2}\xi \Big)   \\
			& + \mathcal{M}	\phi_1'	\psi'\sin^{2}\xi \cos^{2}\xi \Big(\cos\theta_{1}
			+2\hat{\gamma}_{1}\hat{\gamma}_{3}\sin^{2}\theta_{1}\sin^{2}\theta_{2}
			\sin^{2}\xi\cos^{2}\xi \Big)  \\
			& -\mathcal{M}\phi_2'	\psi'\sin^{2}\xi \cos^{2}\xi \Big(\cos\theta_{2}
			-2\hat{\gamma}_{2}\hat{\gamma}_{3}\sin^{2}\theta_{1}\sin^{2}\theta_{2}
			\sin^{2}\xi\cos^{2}\xi \Big)  \\
			& - \frac{1}{2}\mathcal{M}\phi_1'	\phi_2'\sin^{2}\xi\cos^{2}\xi \Big(
			\cos\theta_{1} \cos\theta_{2} -4\hat{\gamma}_{1}\hat{\gamma}_{2}
			\sin^{2}\theta_{1}\sin^{2}\theta_{2}
			\sin^{2}\xi\cos^{2}\xi \Big) \Bigg]
\end{split}		\label{act:BDa}	\\
\begin{split}
&=~ -\frac{1}{2} \Bigg[\frac{1}{4} \dot{t}^{2}-\dot{\xi}^{2}
-\frac{1}{4}\qty(\dot{\theta}_{1}^{2} \cos^{2}\xi +\dot{\theta}_{2}^{2}
\sin^{2}\xi ) +\frac{\mathcal{M}\alpha_{2}^{2}}{4}\cos^{2}\xi
\Big( \sin^{2}\theta_{1}+\sin^{2}\xi \cos^{2}\theta_{1}   \\
& +4\hat{\gamma}_{1}^{2}\sin^{2}\theta_{1}\sin^{2}\theta_{2}
\sin^{4}\xi\cos^{2}\xi \Big)+\frac{\mathcal{M}\alpha_{4}^{2}}{4}\sin^{2}\xi
\Big( \sin^{2}\theta_{2}+\cos^{2}\xi\cos^{2}\theta_{2}+4\hat{\gamma}_{2}^{2} 
\sin^{2}\theta_{1}    \\
& \times \sin^{2}\theta_{2}\sin^{2}\xi\cos^{4}\xi \Big) +\mathcal{M}\alpha_{6}^{2}
\sin^{2}\xi \cos^{2}\xi \Big( 1+\hat{\gamma}_{3}^{2} \sin^{2}\theta_{1}\sin^{2}
\theta_{2}\sin^{2}\xi\cos^{2}\xi \Big)   \\
& + \mathcal{M}\alpha_{2}\alpha_{6}\sin^{2}\xi \cos^{2}\xi \Big(\cos\theta_{1}
+2\hat{\gamma}_{1}\hat{\gamma}_{3}\sin^{2}\theta_{1}\sin^{2}\theta_{2}
\sin^{2}\xi\cos^{2}\xi \Big)  \\
& -\mathcal{M}\alpha_{4}\alpha_{6}\sin^{2}\xi \cos^{2}\xi \Big(\cos\theta_{2}
-2\hat{\gamma}_{2}\hat{\gamma}_{3}\sin^{2}\theta_{1}\sin^{2}\theta_{2}
\sin^{2}\xi\cos^{2}\xi \Big)  \\
& - \frac{1}{2}\mathcal{M}\alpha_{2}\alpha_{4}\sin^{2}\xi\cos^{2}\xi \Big(
\cos\theta_{1} \cos\theta_{2} -4\hat{\gamma}_{1}\hat{\gamma}_{2}
\sin^{2}\theta_{1}\sin^{2}\theta_{2}
\sin^{2}\xi\cos^{2}\xi \Big) \Bigg]  \, .
\end{split}  \label{act:BD} 
\end{alignat}
\end{subequations}
%%%%

\subsubsection{Analytical results}\label{ana:BD}
We begin our analysis by first finding the equations of motion (eom) corresponding to the
non-isometry directions $\theta_{1}$, $\theta_{2}$ and $\xi$ from the Lagrangian density 
(\ref{act:BD}). The results may formally be written as
%%%%
\begin{subequations}\label{eom:BD}
\begin{alignat}{3}
\begin{split}
8\ddot{\theta}_{1} \, \cos^{2}\xi -8\dot{\theta}_{1}\dot{\xi}\, \sin 2\xi
-\partial_{\theta_{1}}\mathcal{M} \cdot T^{(1)}_{\theta_{1}}+\mathcal{M}
\cdot T^{(2)}_{\theta_{1}}  &= 0 \, ,
\end{split}  \label{t1:BD}\\[6pt]
\begin{split}
8\ddot{\theta}_{2} \, \sin^{2}\xi +8\dot{\theta}_{2}\dot{\xi}\, \sin 2\xi
-\partial_{\theta_{2}}\mathcal{M} \cdot T^{(1)}_{\theta_{2}}-\mathcal{M}
\cdot T^{(2)}_{\theta_{2}}  &= 0 \, ,
\end{split}  \label{t2:BD}\\[6pt]
\begin{split}
32 \ddot{\xi}-4\sin2\xi \qty(\dot{\theta}_{2}^{2}-\dot{\theta}_{1}^{2})
-\partial_{\xi}\mathcal{M} \cdot T^{(1)}_{\xi}-\mathcal{M}
\cdot T^{(2)}_{\xi}  &= 0  \, ,
\end{split}  \label{xi:BD}
\end{alignat}
\end{subequations}
%%%%
where 
%%%%
\begin{subequations}\label{derM:BD}
\begin{alignat}{3}
\begin{split}
\partial_{\theta_{1}}\mathcal{M} &= -2 \mathcal{M}^{2}\sin^{2}\xi \cos^{2}\xi
\sin\theta_{1} \Big( \cos\theta_{1}\cos^{2}\xi \qty(4\hat{\gamma}_{2}^{2}
+\hat{\gamma}_{3}^{2}+4\hat{\gamma}_{2}\hat{\gamma}_{3}\cos\theta_{2})
\\
&\quad +2\hat{\gamma}_{1}\hat{\gamma}_{3}\sin^{2}\theta_{2}\sin^{2}\xi \Big) \, ,
\end{split}  \\[3pt]
\begin{split}
 \partial_{\theta_{2}}\mathcal{M} &= -2 \mathcal{M}^{2}\sin^{2}\xi \cos^{2}\xi
\sin\theta_{2} \Big( \cos\theta_{2}\sin^{2}\xi \qty(4\hat{\gamma}_{1}^{2}
+\hat{\gamma}_{3}^{2}-4\hat{\gamma}_{1}\hat{\gamma}_{3}\cos\theta_{1})
\\
&\quad -2\hat{\gamma}_{2}\hat{\gamma}_{3}\sin^{2}\theta_{1}\cos^{2}\xi \Big) \, ,
\end{split}  \\[3pt]
\begin{split}
\partial_{\xi}\mathcal{M} &= -\mathcal{M}^{2}\sin2\xi \Big[ \qty(2\hat{\gamma}_{2}
+\hat{\gamma}_{3}\cos\theta_{2})\cos^{4}\xi \sin^{2}\theta_{1}+\hat{\gamma}_{3}
\cos^{2}\xi \Big\{2\qty(-4\hat{\gamma}_{1}+\hat{\gamma}_{3}\cos\theta_{1})  
\\
&\quad \times \cos\theta_{1}\sin^{2}\theta_{2}\sin^{2}\xi +\sin^{2}\theta_{1} 
\qty(\hat{\gamma}_{3}\sin^{2}\theta_{2}-2\cos\theta_{2}\qty(4\hat{\gamma}_{2}
+\hat{\gamma}_{3}\cos\theta_{2})\sin^{2}\xi) \Big\}
\\
&\quad -\sin^{2}\theta_{1} \qty(2\hat{\gamma}_{2}^{2}\sin^{2}2\xi +
\hat{\gamma}_{3}^{2}\sin^{2}\xi \sin^{2}\theta_{2})
\\
&\quad +\sin^{2}\theta_{2}\qty(2\hat{\gamma}_{1}^{2}\sin^{2}2\xi -\sin^{4}\xi 
\qty(-2\hat{\gamma}_{1}+\hat{\gamma}_{3}\cos\theta_{1})^{2}) \Big]  \, .
\end{split}  
\end{alignat}
\end{subequations}
%%%%

The detailed expressions for the coefficients $T^{(j)}_{i}$ ($j=1,2$, $i=\theta_{1},\theta_{2}
,\xi$) that appear in the above eqs.(\ref{t1:BD})-(\ref{xi:BD}) are provided in the Appendix 
\ref{cofT:BD}.

In the next step, we use (\ref{act:BDa}) to calculate the conjugate momenta\footnote{Here we
use the standard definition of the conjugate momenta as $P_{q}=\partial L_{P}/\partial
\dot{q_{i}}$, where $q_{i}$ are the canonical coordinates.} associated with the isometry
coordinates as
%%%%
\begin{equation}\label{conmom:BD}
E \equiv \pdv{L_{P}}{\dot{t}} = -\frac{\dot{t}}{4} \, , \qquad P_{\Phi_{i}}
\equiv \pdv{L_{P}}{\dot{\Phi}_{i}} = 0 \, ,
\end{equation}
%%%%
where $\Phi_{i}=\{ \phi_{1},\phi_{2},\psi \}$.

From (\ref{conmom:BD}) it is clear that the requirement of the conservation of the momenta,
$J_{i}$,\footnote{Here we define the
charge as
%%%%
\begin{equation}\label{chgJ:BD}
J_{i} = \frac{1}{2\pi\alpha'} \int_{0}^{2\pi} \dd\sigma P_{i} \, ,
\end{equation}
%%%%
where $P_{i}$ are the conjugate momenta.}
given as
%%%%
\begin{equation}\label{conchg:BD}
\partial_{\tau}J_{i} = 0 \, ,
\end{equation}
%%%%
is trivially satisfied. Moreover, the conservation of energy ($\partial_{\tau}E=0$) requires us to
choose the gauge $t=\tau$.
\iffalse
leads to the following constraints:
%%%%
\begin{equation}\label{const:BD}
\ddot{t} =0 \, , \qquad \Pi_{\theta_{1}} :=\dot{\theta}_{1}=0 \, ,
\qquad  \Pi_{\theta_{2}} :=\dot{\theta}_{2}=0 \, , \qquad \Pi_{\xi}
:=\dot{\xi}=0 \, .
\end{equation}
%%%%
\fi

In addition, using (\ref{Hamil}), (\ref{ansatz:BD}) and the eoms (\ref{eom:BD}) it is easy to
check that\footnote{This is an easy but lengthy calculation. Here we avoid writing this very
long expression in order to avoid cluttering.} the Hamiltonian of the system is indeed conserved 
\emph{on-shell}, namely
%%%%
\begin{equation}\label{dervir:BD}
\partial_{\tau}T_{\tau\tau} =0 \, ,
\end{equation}
%%%%
which satisfies the consistency requirement of the Virasoro constraints. On the other hand, using
(\ref{EM:gen}) and (\ref{ansatz:BD}) we observe that the non-diagonal component of the
energy-momentum tensor ($T_{\tau\sigma}$) is also conserved trivially, namely $\displaystyle 
\partial_{\tau}T_{\tau\sigma}=0$.

The dynamics of the string is described by the eoms (\ref{t1:BD})-(\ref{xi:BD}). In order to study
the string configuration methodically, we first choose the $\theta_{2}$ invariant plane in the phase
space given by
%%%%
\begin{equation}
\label{IP1:BD}
\theta_{2}\sim 0 \, , \qquad \quad \Pi_{\theta_{2}}:=\dot{\theta}_{2}=0 \,.
\end{equation}
%%%%

Notice that, the above choice (\ref{IP1:BD}) trivially satisfies the $\theta_{2}$ eom (\ref{t2:BD}).
On the other hand, the remaining two eoms (\ref{t2:BD}) and (\ref{xi:BD}) become
%%%%
\begin{subequations}
\begin{alignat}{2}
\begin{split}
8\ddot{\theta}_{1} \, \cos^{2}\xi -8\dot{\theta}_{1}\dot{\xi}\, \sin 2\xi
-\widetilde{\partial_{\theta_{1}}\mathcal{M}}\cdot
\widetilde{T^{(1)}_{\theta_{1}}} + \widetilde{\mathcal{M}}\cdot
\widetilde{T^{(2)}_{\theta_{1}}} &= 0  \, ,
\end{split}  \label{t12:BD} \\[5pt]
\begin{split}
32 \ddot{\xi}+4\sin2\xi \dot{\theta}_{1}^{2}
-\widetilde{\partial_{\xi}\mathcal{M}} \cdot \widetilde{T^{(1)}_{\xi}}-
\widetilde{\mathcal{M}}\cdot \widetilde{T^{(2)}_{\xi}}  &= 0  \, ,
\end{split}  \label{xi2:BD}
\end{alignat}
\end{subequations}
%%%%
where 
%%%%
\begin{subequations}
\begin{alignat}{7}
\begin{split}
\widetilde{\mathcal{M}} &= \qty(1+\qty(2\hat{\gamma}_{2}
+\hat{\gamma}_{3})^{2}\sin^{2}\theta_{1}\sin^{2}\xi
\cos^{4}\xi)^{-1}  \, ,
\end{split}   \\[5pt]
\begin{split}
\widetilde{\partial_{\theta_{1}}\mathcal{M}} &= -\frac{\qty(2
\hat{\gamma}_{2}+\hat{\gamma}_{3})^{2}\sin^{2}\xi
\cos^{4}\xi \sin 2\theta_{1}}{\qty(1+\qty(2\hat{\gamma}_{2}
+\hat{\gamma}_{3})^{2}\sin^{2}\theta_{1}\sin^{2}\xi
\cos^{4}\xi)^{2}}  \, ,
\end{split} \\[5pt]
\begin{split}
\widetilde{\partial_{\xi}\mathcal{M}} &= -\frac{\qty(2
\hat{\gamma}_{2}+\hat{\gamma}_{3})^{2}\qty(-1+3\cos 2\xi)
\sin\xi \cos^{3}\xi \sin^{2}\theta_{1}}{\qty(1+\qty(2\hat{\gamma}_{2}
+\hat{\gamma}_{3})^{2}\sin^{2}\theta_{1}\sin^{2}\xi
\cos^{4}\xi)^{2}}  \, ,
\end{split} \\[5pt]
\begin{split}
\widetilde{T^{(1)}_{\theta_{1}}} &= -4\cos^{2}\xi \Big[\alpha_{2}^{2}
\sin^{2}\theta_{1}+\alpha_{6}^{2}\sin^{2}\xi +\sin^{2}\xi
\Big(\qty(\alpha_{4}-\alpha_{2}\cos\theta_{1})^{2}      \\
&\quad +4\alpha_{6}\qty(\alpha_{2}\cos\theta_{1}-\alpha_{4}) \Big)
\Big]  \, ,
\end{split} \\[5pt]
\begin{split}
\widetilde{T^{(2)}_{\theta_{1}}} &= \alpha_{2}^{2}\sin 2\theta_{1}
\qty(4\cos^{2}\xi -\sin^{2}2\xi) +2\alpha_{2}\sin^{2}2\xi \sin\theta_{1}
\qty(\alpha_{4}-2\alpha_{6})  \, ,
\end{split} \\[5pt]
\begin{split}
\widetilde{T^{(1)}_{\xi}} &= 4\sin^{2}\xi \cos^{2}\xi \qty(-\alpha_{2}^{2}
\cos^{2}\theta_{1}+2\alpha_{2}\alpha_{4}\cos\theta_{1}-\alpha_{4}^{2}
-4\alpha_{2}\alpha_{6}\cos\theta_{1}+4\alpha_{4}\alpha_{6})  \\
&\quad -4\qty(\alpha_{2}^{2}\sin^{2}\theta_{1}\cos^{2}\xi +\alpha_{6}^{2}
\sin^{2}2\xi)  \, ,
\end{split} \\[5pt]
\begin{split}
\widetilde{T^{(2)}_{\xi}} &= -2 \Big[\alpha_{2}^{2}\qty(\sin4\xi \cos^{2}
\theta_{1} -2\sin^{2}\theta_{1}\sin2\xi)+\sin 4\xi \Big(2\alpha_{2}
\cos\theta_{1}\qty(2\alpha_{6}-\alpha_{4}) +\qty(\alpha_{4}-2\alpha_{6})^{2}
\Big) \Big] \, .
\end{split} 
\end{alignat}
\end{subequations}
%%%%

In the next step, in order to utilize the Kovacic's algorithm to the string configuration in the reduced
phase-space described by (\ref{t12:BD}) and (\ref{xi2:BD}), we make the choice
%%%%
\begin{align}
\label{IP2:BD}
\theta_{1}  \sim 0 \, , \qquad \Pi_{\theta_{1}} \equiv \dot{\theta}_{1} \sim 0 \, .
\end{align}
%%%%

Eq.(\ref{IP2:BD}) indeed satisfies (\ref{t12:BD}), and the remaining eom (\ref{xi2:BD})
can be recast in the form
%%%%
\begin{equation}
\label{xi3:BD}
\ddot{\xi}+\mathcal{A}_{\text{BD}} \sin4\xi = 0 \, ,
\end{equation}
%%%%
where
%%%%
\begin{equation}
\label{ABD:BD}
\mathcal{A}_{\text{BD}} = \frac{1}{16}\qty[\alpha_{2}^{2}
+2\alpha_{2}\qty(2\alpha_{6}-\alpha_{4})+\qty(\alpha_{4}
-2\alpha_{6})^{2}] \, .
\end{equation}
%%%%

In order to proceed farther, we consider infinitesimal fluctuation ($\eta$) around the $\theta_{1}$
invariant plane in the phase space. Considering terms only upto $\order{\eta}$, we may re-express
(\ref{t12:BD}) as
%%%%
\begin{align}\label{etaNVE:BD}
\begin{split}
& 8\ddot{\eta} \cos^{2}\bar{\xi}-8\dot{\bar{\xi}}\sin2\bar{\xi}\, \dot{\eta}
+\Big( 8\alpha_{2}\cos^{2}\bar{\xi}\qty(\alpha_{2}\cos^{2}\bar{\xi}+\alpha_{4}
\qty(\alpha_{4}-2\alpha_{6})\sin^{2}\bar{\xi})    \\
&\quad -8 \qty(2\hat{\gamma}_{2}+\hat{\gamma}_{3})^{2}\sin^{4}\bar{\xi}
\cos^{6}\bar{\xi}  \qty(4\alpha_{6}^{2}+\qty(\alpha_{2}-\alpha_{4})^{2}+
4\alpha_{6}\qty(\alpha_{2}-\alpha_{4})) \Big)\eta =0 \, ,
\end{split}
\end{align}
%%%%
where $\bar{\xi}$ is the solution to (\ref{xi3:BD}).

In order to study (\ref{etaNVE:BD}), we make the change in variable as
%%%%
\begin{equation}
\label{varz:BD}
\cos\bar{\xi} = z \, .
\end{equation} 
%%%%

Using (\ref{varz:BD}) we can convert (\ref{etaNVE:BD}) to a second order linear homogeneous
differential equation, known as the Lam\'{e} equation \cite{Basu:2011fw}, as
%%%%
\begin{equation}
\label{zNVE:BD}
\eta''(z)+B(z)\eta'(z)+A(z)\eta(z) =0 \, ,
\end{equation}
%%%%
where
%%%%
\begin{subequations}
\begin{alignat}{3}
\begin{split}
B(z) &= \frac{f'(z)}{2f(z)}+\frac{2}{z}  \, ,
\end{split}  \label{B:BD} \\[5pt]
\begin{split}
f(z) &= \dot{\bar{\xi}}^{2}\sin^{2}\bar{\xi} = \qty(E+
\frac{\mathcal{A}_{\text{BD}}}{2}\qty(8z^{4}-8z^{2}+1))
\qty(1-z^{2})  \, ,
\end{split}  \label{f:BD}  \\[5pt]
\begin{split}
A(z) &=~ \Big(\alpha_{2}\qty(\alpha_{2}z^{2}+\qty(\alpha_{4}-2\alpha_{6})
\qty(1-z^{2}))
\\
&~-\qty(2\gamma_{2}+\gamma_{3})^{2}\qty(4\alpha_{6}^{2}+
\qty(\alpha_{2}-\alpha_{4})\qty(\alpha_{2}-\alpha_{4}+4\alpha_{6}))
z^{4}\qty(1-z^{2})^{2}\Big) \cdot \frac{1}{f(z)}   \, .
\end{split}  \label{A:BD}
\end{alignat}
\end{subequations}
%%%%
In our subsequent analysis we choose the string energy $E=1$ in (\ref{f:BD}).

We can farther express (\ref{zNVE:BD}) in the Schr\"{o}dinger form (\ref{Kov:Fnl}) by using
the change in variable (\ref{Kov:Var}). The result may formally be written as
%%%%
\begin{equation}
\label{formsch:BD}
\omega'(z)+\omega^{2}(z)=\frac{2B'(z)+B^{2}(z)-4A(z)}{4}
\equiv \mathcal{V}_{\text{BD}}(z) \, ,
\end{equation}
%%%%
where the potential is given by
%%%%
\begin{align}\label{formV:BD}
\begin{split}
\mathcal{V}_{\text{BD}} &=~
\Bigg\{ 8\alpha_{2}\qty(z^{2}-1) \Big( \alpha_{2}z^{2} -\qty(z^{2}-1) \qty(\alpha_{2}
-2\alpha_{6})-z^{2}\qty(z^{2}-1)^{2} \qty(\alpha_{2}-\alpha_{4}+2\alpha_{6})^{2}
\qty(2\hat{\gamma}_{2}+\hat{\gamma}_{3})^{2}\Big)
\\
&~ \times \Big(2+\qty(1-8z^{2}+8z^{4})\mathcal{A}_{\text{BD}}\Big)
+z^{-2} \Big( -4+6z^{2}+\qty(-2+27z^{2}-64z^{4}+40z^{6})
\mathcal{A}_{\text{BD}} \Big)^{2}
\\
&~ -\Bigg[ 2z^{-2} \Big\{ 4\qty(2-3z^{2}+3z^{4})+4\qty(2-15z^{2}+11z^{4}
+4z^{6})\mathcal{A}_{\text{BD}}
\\
&~ +\qty( 2-27 z^{2}+211 z^{4}-632 z^{6}+1024 z^{8}-896 z^{10}+320 z^{12})
\mathcal{A}_{\text{BD}} \Big\} \Bigg] \Bigg\}
\\
&~ \times \frac{1}{4 \left(z^2-1\right)^2 \Big(\left(8 z^4-8 z^2+1\right) 
\mathcal{A}_{\text{BD}}+2\Big)^2} \, .
\end{split}
\end{align}
%%%%

In order to find the solution to (\ref{formsch:BD}), we first notice that the value of $\xi$
cannot be zero since this implies that one of the two-spheres in the $CP^{3}$ space in
(\ref{met:BD}) vanishes. This restricts our analysis to a particular subspace of the $CP^{3}$ space. 
However, since we want to take into consideration the entire metric space, we exclude this possibility. 
Hence, $0 \leq |z| < 1$. This argument allows us to expand the potential $\mathcal{V}_{\text{BD}}$
in $z$. In the leading order in $z$, (\ref{formsch:BD}) is found to have the form
%%%%
\begin{equation}
\label{redSch:BD}
\omega'(z)+\omega^{2}(z) = \widetilde{C}_{1} \, . 
\end{equation}
%%%%
where 
%%%%
\begin{equation}
\widetilde{C}_{1} = ~ - \frac{4 \alpha_{2} \alpha_{4}-8 \alpha_{2}
\alpha_{6}+27 \mathcal{A}_{\text{BD}}+6}{2 \mathcal{A}_{\text{BD}}+4} \, .
\end{equation}
%%%%

The solution to (\ref{redSch:BD}) is found to be
%%%%
\begin{equation}
\label{solSch:BD}
\omega(z) = \sqrt{\widetilde{C}_{1}}~ \tanh \Big[ \sqrt{\widetilde{C}_{1}}
~\qty(z+\mathsf{C}_{1}) \Big] \, ,
\end{equation}
%%%%
where $\mathsf{C}_{1}$ is an arbitrary integration constant.

Now from (\ref{formV:BD}) we observe that the potential has poles of order $2$ at the
following values of $z$:
%%%%
\begin{equation}\label{poles:BD}
z = \pm 1 \, , \quad z = \pm \frac{1}{2}\sqrt{2-\frac{\sqrt{2}
\sqrt{\mathcal{A}_{\text{BD}}\qty(\mathcal{A}_{\text{BD}}-2)}}
{\mathcal{A}_{\text{BD}}}} \, , \quad z = \pm \frac{1}{2}\sqrt{2+
\frac{\sqrt{2}\sqrt{\mathcal{A}_{\text{BD}}
\qty(\mathcal{A}_{\text{BD}}-2)}}{\mathcal{A}_{\text{BD}}}} \, .
\end{equation}
%%%%
On the other hand, from (\ref{formV:BD}) the \emph{order at infinity} of
$\mathcal{V}_{\text{BD}}$ is found to be $2$. Thus 
$\mathcal{V}_{\text{BD}}$ satisfies the condition \textbf{Cd.}$(iii)$ of the Kovacic's classification 
discussed in Appendix \ref{Kova}. On top of that, for small values of $z$ the solution
(\ref{solSch:BD}) is indeed a polynomial of degree $1$. This matches one of the integrability
criteria as put forward by the Kovacic's algorithm discussed in Appendix \ref{Kova}.

In order to support our analytic result, below we numerically check the non-chaotic dynamics
of the propagating string.

\subsubsection{Numerical results}\label{num:BD}
In our numerical analysis, we use the ansatz (\ref{ansatz:BD}) together with the choice
$\alpha_{i}=1$ ($i=2,4,6$) of the winding numbers. The resulting Hamilton's equations of
motion can be written as
%%%%
\begin{subequations}\label{Hamil:BD}
\begin{alignat}{4}
\begin{split}
\dot{\theta}_{1} =& 4p_{\theta_{1}}\sec^{2}\xi  \, ,
\end{split}  \label{Hamilth:BD}  \\
\begin{split}
\dot{\xi} =& p_{\xi}  \, ,
\end{split}  \label{Hamilxi:BD}  \\
\begin{split}
\dot{\theta}_{1} =& \frac{\cos^2\xi\sin\theta_{1}
\Big(-4\cos\theta_{1}\cos^2\xi+4\sin^2\xi+(2\hat{\gamma}_2
+\hat{\gamma}_3)^2\cos^4{(\theta_1/2)}\sin^4(2\xi)\Big)}
{16\Big(1+(2\hat{\gamma}_2+\hat{\gamma}_3)^2\cos^4\xi
\sin^2\theta_{1}\sin^2\xi \Big)^2}
\end{split}  \label{Hamilpth:BD}  \\
\begin{split}
\dot{p_{\xi}}=& \frac{\mathcal{N}_{1}}
{128\Big(1+(2\hat{\gamma}_2+\hat{\gamma}_3)^2\cos^4\xi
\sin^2\theta_{1}\sin^2\xi \Big)^2}
\end{split}  \label{Hamilpxi:BD} 
\end{alignat}
\end{subequations}
%%%%
where
%%%%
\begin{align}
\mathcal{N}_1=&\hspace{1mm}16 (2\hat{\gamma}_2+\hat{\gamma}_3)^2
\cos^5\xi (-1+3\cos(2\xi))\sin^2\theta_{1}\sin^3\xi+32(2\hat{\gamma}_2
+\hat{\gamma}_3)^2\cos\theta_{1}\cos^5\xi(-1+       \nonumber\\
&3\cos(2\xi))\sin^2\theta_{1}\sin^3\xi+(2\hat{\gamma}_2+\hat{\gamma}_3)^2
\cos^5\xi(-6+2\cos(2\theta_{1})+\cos(2\theta_{1}-2\xi)+     \nonumber\\
&2\cos(2\xi)+\cos(2\theta_{1}+2\xi))\sin^2\theta_{1}
(5\sin\xi-3\sin(3\xi))-8(1+(2\hat{\gamma}_2+\hat{\gamma}_3)^2\cos^4\xi
\times      \nonumber\\
&\sin^2\theta_{1}\sin^2\xi)\sin(4\xi)-16\cos\theta_{1}
(1+(2\hat{\gamma}_2+\hat{\gamma}_3)^2\cos^4\xi\sin^2\theta_{1}\sin^2\xi)
\sin(4\xi)\nonumber\\&+4(1+(2\hat{\gamma}_2+\hat{\gamma}_3)^2\cos^4\xi
\sin^2\theta_{1}\sin^2\xi)(4\sin^2\theta_{1}\sin(2\xi)-2\cos^2\theta_{1}\sin(4\xi))
      \nonumber\\
&-512p^2_{\theta_{1}}\sec^2\xi(1+(2\hat{\gamma}_2+\hat{\gamma}_3)^2
\cos^4\xi\sin^2\theta_{1}\sin^2\xi)^2\tan{\xi}.
\end{align}
%%%%

It must be stressed that, in writing the Hamilton's eoms (\ref{Hamil:BD}), we set
$\theta_{2}=p_{\theta_{2}}=0$.

%%%%
\begin{figure}[t]
\begin{multicols}{2}
    \includegraphics[width=\linewidth]{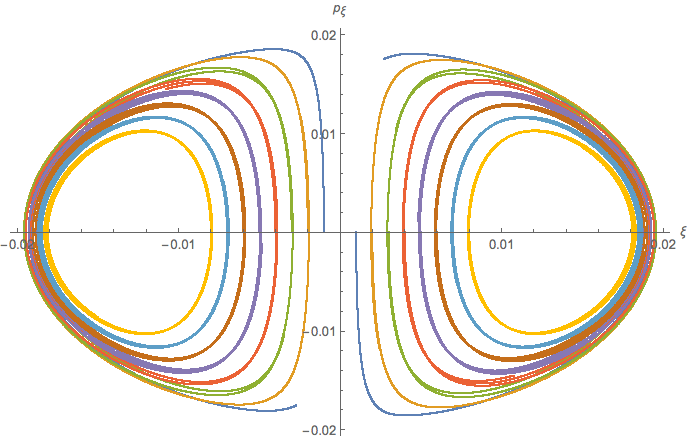}\par 
    \includegraphics[width=\linewidth]{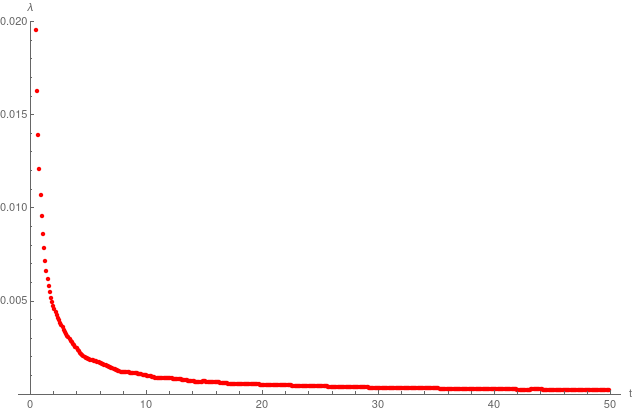}\par 
    \end{multicols}
\begin{multicols}{2}
    \includegraphics[width=\linewidth]{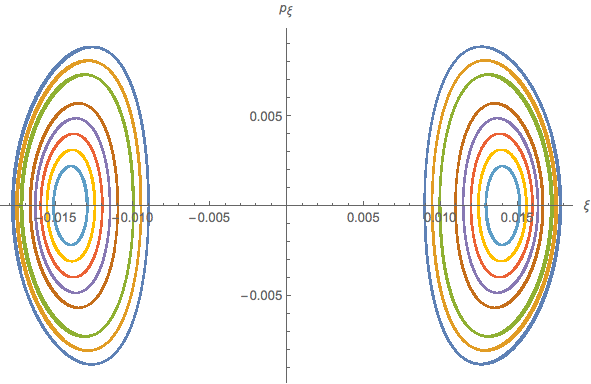}\par
    \includegraphics[width=\linewidth]{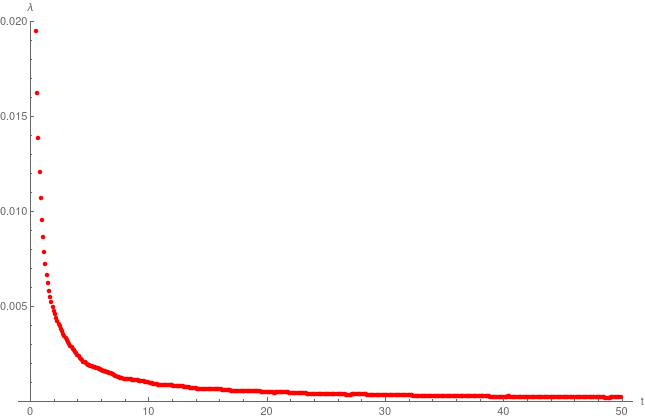}\par
\end{multicols}
\caption{Numerical plots of the Poincar\'{e} sections (\textbf{Left column}) and
Lyapunov exponents (\textbf{Right column}) for $\beta$-deformed ABJM. Here
we set the energy of the string $E_{0}=0.01$. The top plots are for
$\hat{\gamma}_{2}= \hat{\gamma}_{3}=0.01$ and the bottom plots are for
$\hat{\gamma}_{2} =\hat{\gamma}_{3}=0.8$. The Poincar\'{e} sections are
nicely foliated KAM tori and the Lyapunov exponent decays to zero for large time
$t$, indicating non-chaotic dynamics of the string.}
\label{fig1:BD}
\end{figure}
%%%%

In order to obtain the corresponding Poincar\'{e} sections, we solve the Hamiltonian's eoms 
(\ref{Hamilth:BD})-(\ref{Hamilpxi:BD}) subjected to the constraints (\ref{Hamil}) and
(\ref{app:Vir}). These are plotted in the left column of fig.\ref{fig1:BD}. The energy of the
string is fixed at some particular value $E = E_0=0.01$, whereas the values of the YB
deformation parameters are set as $\hat{\gamma}_2=\hat{\gamma}_3=0.01, ~ 0.8$.
In addition, we choose the initial conditions as $\theta_{1}(0)=0$ and $p_\xi(0)=0$. Given
this initial set of data, we generate a random data set for an interval $\xi(0)\in [0, 1]$ which
fixes the corresponding $ p_{\theta_{1}}(0)$ in accordance with that of the constraint
(\ref{Hamil}).

It is important to note that the other deformation parameter $\hat{\gamma}_1$ disappears from
the numerical simulation since we switch off the $\qty{\theta_{2}, p_{\theta_{2}}}$ variables
in the phase space. This is also visible form the Hamilton's eoms
(\ref{Hamilth:BD})-(\ref{Hamilpxi:BD}), which do not depend on the choice of
$ \hat{\gamma}_{1} $.

We plot all these points on the $\lbrace\xi,p_\xi\rbrace$ plane every time the trajectories pass
through $\theta_{1}=0$ hyper-plane. For the present example, the phase space under
consideration is four dimensional, namely it is characterized by the coordinates $\lbrace\theta_{1},
p_{\theta_{1}},\xi,p_\xi \rbrace$. Poincare sections in this case show regular patches indicating
a foliation in the phase space (cf. left column of fig.\ref{fig1:BD}).

In order to calculate the Lyapunonv exponent ($\lambda$), we choose to work with the initial
conditions $E = E_0=0.01$ together with $\lbrace\theta_{1}(0)=0, \xi (0)=0.008,
p_{\theta_{1}}(0)=0.009, p_\xi(0)=0\rbrace$ which are consistent with (\ref{Hamil}). When
$\hat{\gamma}_{2}=\hat{\gamma}_{3}=0.8 $, the initial conditions are set to be
$\lbrace\theta_{1}(0)=0, \xi (0)=0.013, p_{\theta_{1}}(0)=0.007, p_\xi(0)=0\rbrace$ while
we keep the energy to be fixed at $E = E_0=0.01$. With this initial set of data, we study the
dynamical evolution of two nearby orbits in the phase space those have an initial separation
$ \Delta X_{0}=10^{-7} $ (cf. (\ref{app:Lyav}) ). In the process, we generate a zero
Lyapunov exponent at large $t$ as shown in the right column of fig.\ref{fig1:BD}. This observation indeed exhibits a non-chaotic dynamics  of the super string in the phase space. Moreover, we also verified the above conclusion by permitting higher values of the string energy as shown in Fig.\ref{Fig:New1:BD} below.

%%%%
\begin{figure}[t]
	\begin{multicols}{2}
		\includegraphics[width=\linewidth]{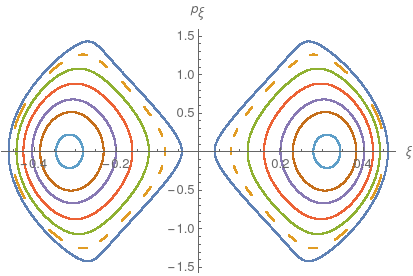}\par 
		\includegraphics[width=\linewidth]{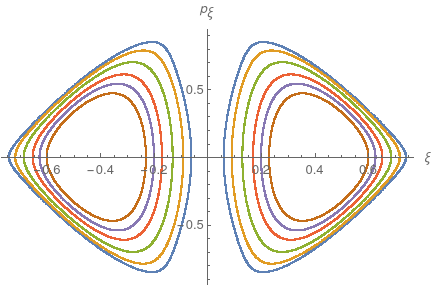}\par 
	\end{multicols}
	\caption{Additional plots of the Poincar\'{e} sections for $\beta$-deformed ABJM. On the left plot we set $\mathbf{E_{0}=1}$, $\hat{\gamma}_{2}= \hat{\gamma}_{3}=0.5$. The plot on right corresponds to  $\mathbf{E_{0}=0.5}$, 
		$\hat{\gamma}_{2}= \hat{\gamma}_{3}=0.1$.}
	\label{Fig:New1:BD}
\end{figure}
%%%%

%%%%%%%%%%%%%%%%%%%%%%%%%%%%%%%%%%
\subsection{Noncommutative ABJM}\label{ABJM:NC}

Noncommutative ABJM corresponds to a gravity dual that is obtained by applying YB deformation
to its AdS$_{4}$ subspace. The corresponding Abelian $ r $-matrix is constructed using the
momenta operators along AdS$_{4}$ \footnote{The form of the $r$-matrix is taken to be
\begin{equation*}
r =~ \mu \, \mathbf{p}_{1} \wedge \mathbf{p}_{2} \, ,
\end{equation*}
where $\mathbf{p}_{1}$ and $\mathbf{p}_{2}$ are the momentum operators along the $x_{1}$
and $x_{2}$ directions, respectively. The $B$-field (\ref{ns:NC}) results in the noncommutativity
$\displaystyle \left[ x_{1},x_{2}\right] \sim \mu$ in the $x_{1}-x_{2}$ plane
\cite{Rado:2020swb}. }. The resulting space-time metric is given by \cite{Rado:2020swb}
%%%%
\begin{align}\label{met:NC}
\begin{split}
\dd s^{2}=&\hspace{1mm}\frac{1}{4}\left(r^{2}\left(-dt^{2}+\mathcal{M}
\left(d x_{1}^{2}+d x_{2}^{2}\right)\right)+\frac{d r^{2}}{r^{2}}\right)
+d s_{CP^{3}}^{2}  \, ,  \\[7pt]
\dd s^{2}_{ CP^3} =&\hspace{1mm}\mathrm{d}\xi^2+\frac{1}{4}
\cos^2{\xi}(\mathrm{d}\theta_1^2+\sin^2{\theta_1}\mathrm{d}\phi_1^2)
+\frac{1}{4}\sin^2{\xi}(\mathrm{d}\theta_2^2+\sin^2{\theta_2}\mathrm{d}
\phi_2^2)\\
& +\bigg(\frac{1}{2}\cos{\theta_1}\mathrm{d}\phi_1-\frac{1}
{2}\cos{\theta_2}\mathrm{d}\phi_2+\mathrm{d}\psi\bigg)^2\sin^2{\xi}\cos^2{\xi} \, ,
\end{split}
\end{align}
%%%%
which is accompanied by a NS-NS two form
%%%%
\begin{align}\label{ns:NC}
B=\frac{\mu \mathcal{M} r^{4}}{4} d x^{1} \wedge d x^{2},\hspace{2mm}\mathcal{M}^{-1}=1+\frac{\mu^{2} r^{4}}{4} \, ,
\end{align}
%%%%
where $t$, $x_1$, $x_2$ and $r$ are the coordinates of $AdS_4$ background and $\mu$ is
the YB deformation parameter. We set $r=1$ for the rest of our analysis.

In the next step, we consider the winding string ansatz of the form
%%%%
\begin{align}\label{ansatz:NC}
t=&\hspace{1mm}t(\tau), \quad \theta_1=\theta_1(\tau),
\quad \theta_2=\theta_2(\tau),\quad \xi=\xi(\tau),\quad
\phi_1=\alpha_2\sigma,\quad \phi_2=\alpha_4\sigma,\nonumber\\ 
\psi=& \hspace{1mm}\alpha_6\sigma,\quad x_1=\alpha_8\sigma,
\quad x_2=\alpha_{10}\sigma  \, ,
\end{align}
%%%%
where $\alpha_{i}$ ($i=2,4,6,8,10$) are the winding numbers. Notice that, with the above ansatz 
(\ref{ansatz:NC}) the contribution of the $B$-field in the Polyakov action (\ref{POL}) vanishes. 

%In our analytical as well as numerical analyses we make the choice $\alpha_{i}=1$
%($i=2,4,6,8,10$) in (\ref{ansatz:NC}).

In the next step, using (\ref{met:NC}) and (\ref{ansatz:NC}), the Polyakov action (\ref{POL})
can be expressed as\footnote{Notice that, the information of the deformation parameter ($\mu$)
is encoded in $\mathcal{M}$ ((\ref{ns:NC})) which in turn modifies the Hamiltonian (\ref{Hamil}).}
%%%%
\begin{subequations}
\begin{alignat}{2}
\begin{split}
L_{P} &=~ -\frac{1}{2}\left[\frac{1}{4}\dot{t}^{2}-\dot{\xi}^{2}-\frac{1}{4}
		\dot{\theta}_{1}^{2}\,\cos^{2}\xi -\frac{1}{4} \dot{\theta}_{2}^{2}\,
		\sin^{2}\xi +\frac{\mathcal{M}}{4}\qty(x_1'^2+x_2'^2)
		\right. \\[5pt]
		& \quad \left. +\frac{\phi_{1}'^2}{4}\cos^{2}\xi
		\qty(\sin^{2}\theta_{1}+\cos^{2}\theta_{1}\sin^{2}\xi)
		+\frac{\phi_{2}'^2}{4}\sin^{2}\xi 
		\qty(\sin^{2}\theta_{2}+\cos^{2}\theta_{2}\cos^{2}\xi)\right. \\[5pt]
		&\quad \left.+\sin^{2}\xi \cos^{2}\xi \Big\{\psi'^2
		+\phi_{1}'\psi'\cos\theta_{1}-\phi_{2}'\psi'\cos\theta_{2}
		-\frac{1}{2}\phi_{1}'\phi_{2}'\cos\theta_{1}\cos\theta_{2}\Big\}\right]
\end{split}  	\label{act:NCa}	\\[5pt]
\begin{split}
&=~ -\frac{1}{2}\left[\frac{1}{4}\dot{t}^{2}-\dot{\xi}^{2}-\frac{1}{4}
\dot{\theta}_{1}^{2}\,\cos^{2}\xi -\frac{1}{4} \dot{\theta}_{2}^{2}\,
\sin^{2}\xi +\frac{\mathcal{M}}{4}\qty(\alpha_{8}^{2}+\alpha_{10}^{2})
\right. \\[5pt]
& \quad \left. +\frac{\alpha_{2}^{2}}{4}\cos^{2}\xi
\qty(\sin^{2}\theta_{1}+\cos^{2}\theta_{1}\sin^{2}\xi)
+\frac{\alpha_{4}^{2}}{4}\sin^{2}\xi 
\qty(\sin^{2}\theta_{2}+\cos^{2}\theta_{2}\cos^{2}\xi)\right. \\[5pt]
&\quad \left.+\sin^{2}\xi \cos^{2}\xi \Big\{\alpha_{6}^{2}
+\alpha_{2}\alpha_{6}\cos\theta_{1}-\alpha_{4}\alpha_{6}\cos\theta_{2}
-\frac{1}{2}\alpha_{2}\alpha_{4}\cos\theta_{1}\cos\theta_{2}\Big\}\right] \; .
\end{split} \label{act:NC}
\end{alignat}
\end{subequations}
%%%%

\subsubsection{Analytical results}\label{ana:NC}
The equations of motion corresponding to the non-isometry coordinates $\theta_{1}$,
$\theta_{2}$ and $\xi$ can be computed from (\ref{act:NC}) as
%%%%
\begin{subequations}\label{eqmain:NC}
\begin{alignat}{2}
\begin{split}
& \ddot{\theta}_{1}-2\tan\xi \, \dot{\xi}\dot{\theta}_{1}+\alpha_{2}
\sin\theta_{1}\Big( \alpha_{2}\cos^{2}\xi \cos\theta_{1}+
\qty(\alpha_{4}\cos\theta_{2}-2\alpha_{6})\sin^{2}\xi \Big) =0 \, ,
\end{split}  \label{t1:NC} \\[6pt]
\begin{split}
& \ddot{\theta}_{2}+2\cot\xi \, \dot{\xi}\dot{\theta}_{2}+\alpha_{4}
\sin\theta_{2}\Big( \alpha_{4}\sin^{2}\xi \cos\theta_{2}+
\qty(\alpha_{2}\cos\theta_{1}+2\alpha_{6})\cos^{2}\xi \Big) =0 \, ,
\end{split}  \label{t2:NC} \\[6pt]
\begin{split}
& 8\ddot{\xi}+\sin 2\xi \qty(\dot{\theta}_{1}^{2}-\dot{\theta}_{2}^{2})
+2\alpha_{6}^{2} \sin 4\xi -2\alpha_{4}\alpha_{6}\cos\theta_{2}
\sin 4\xi   \\
&-\alpha_{2}\cos\theta_{1}\sin 4\xi \qty(\alpha_{4}\cos\theta_{2}
-2\alpha_{6})+\alpha_{2}^{2}\qty(-\sin^{2}\theta_{1}\sin 2\xi
+\frac{1}{2}\cos^{2}\theta_{1}\sin 4\xi) \\
&+\alpha_{4}^{2}\qty(\sin^{2}\theta_{2}\sin 2\xi +\frac{1}{2}\cos^{2}
\theta_{2}\sin 4\xi) =0 \, .
\end{split} \label{xi:NC}
\end{alignat}
\end{subequations}
%%%%

The conjugate momenta corresponding to the coordinates $\{t, \Phi_{i}\}$
with $(i=\phi_{1},\phi_{2},\psi,x_{1},x_{2})$ can be computed as
%%%%
\begin{equation}\label{conmom:NC}
E \equiv \frac{\partial L_{P}}{\partial\dot{t}} =-\frac{\dot{t}}{4} \, ,
\qquad P_{\Phi_{i}} \equiv \frac{\partial L_{P}}{\partial\dot{\Phi_{i}}}
=0 \, .
\end{equation} 
%%%%

Using (\ref{chgJ:BD}) it is trivial to check that the corresponding charges are indeed conserved.
%%%%
\begin{equation}\label{ConChg:NC}
\partial_{\tau}E=0 \, \quad (\text{in $t=\tau$ gauge}) \, ,
\qquad \partial_{\tau} P_{\Phi_{i}}=0  \, .
\end{equation}
%%%%

Next, following the same line of arguments as in the previous Section \ref{ana:BD}
(cf. eq.(\ref{dervir:BD})), we may easily verify that the energy-momentum tensor
satisfies the Virasoro consistency conditions
%%%%
\begin{align}\label{dervir:NC}
\begin{split}
\partial_{\tau}T_{\tau\tau} &=~0  \, , \quad   \text{on-shell} \, ,
\\[5pt]
\partial_{\tau}T_{\tau\sigma} &=~0  \, , \quad \text{trivially} \, .
\end{split}
\end{align}

The string configuration is described by the three equations of motion (\ref{t1:NC})-(\ref{xi:NC}).
In order to study this configuration systematically, we choose the following invariant plane in the
phase space:
%%%%
\begin{align}
\label{IP1:NC}
\theta_{2} & \sim 0 \, , & \Pi_{\theta_{2}} &\equiv \dot{\theta}_{2} \sim 0 \, .
\end{align}
%%%%
Notice that, the choice (\ref{IP1:NC}) automatically satisfies the $\theta_{2}$ eom (\ref{t2:NC}).
The eoms corresponding to $\theta_{1}$ and $\xi$ then reduce to
%%%%
\begin{subequations}
\begin{alignat}{2}
\begin{split}
&\ddot{\theta}_{1}-2\dot{\xi}\dot{\theta}_{1}\tan\xi +\alpha_{2}\sin\theta_{1}
\qty[\alpha_{2}\cos\theta_{1}\cos^{2}\xi +\sin^{2}\xi \qty(\alpha_{4}-2
\alpha_{6})] =0  \, ,
\end{split}  \label{t12:NC}  \\[6pt]
\begin{split}
&8\ddot{\xi} +\sin 2\xi \, \dot{\theta}_{1}^{2}+\sin 4\xi \Big[2\alpha_{6}^{2}
+\frac{\alpha_{4}^{2}}{2} -2\alpha_{4}\alpha_{6}-\alpha_{2}\cos\theta_{1}
\qty(\alpha_{4}-2\alpha_{6})\Big]  \\
& \qquad\qquad +\alpha_{2}^{2} \Big[ -\sin^{2}\theta_{1}\sin 2\xi
+\frac{1}{2} \sin 4\xi \cos^{2}\theta_{1}\Big] = 0 \, .
\end{split} \label{xi2:NC}
\end{alignat}
\end{subequations}
%%%%

In the next step, in order to utilize the Kovacic's algorithm to the string configuration in the reduced
phase-space described by (\ref{t12:NC}) and (\ref{xi2:NC}), we make the choice
%%%%
\begin{align}
\label{IP2:NC}
\theta_{1} & \sim 0 \, , & \Pi_{\theta_{1}} &\equiv \dot{\theta}_{1} \sim 0 \, .
\end{align}
%%%%

This choice satisfies (\ref{t12:NC}) trivially, and we are left with the following eom:
%%%%
\begin{equation}
\label{xi3:NC}
\ddot{\xi} +\mathcal{A}_{\text{NC}} \sin 4\xi = 0 \, ,
\end{equation}
%%%%
where
%%%%
\begin{equation}
\label{ANC:NC}
\mathcal{A}_{\text{NC}} = \frac{1}{8} \qty[2\alpha_{6}\qty(\alpha_{6}
-\alpha_{4})-\alpha_{2}\qty(\alpha_{4}-2\alpha_{6})+\frac{1}{2}\qty(
\alpha_{2}^{2}+\alpha_{4}^{2})] \, .
\end{equation}
%%%%

We now consider small fluctuations ($\eta$) around the invariant plane $\theta_{1}$. This results
the normal variational equation (NVE) of the form
%%%%
\begin{equation}\label{etaNVE:NC}
\ddot{\eta} - 2 \dot{\bar{\xi}}\tan\bar{\xi} \; \dot{\eta} +\alpha_{2}
\qty[ \alpha_{2}\cos^{2}\bar{\xi} +\qty(\alpha_{4}-2\alpha_{6})
\sin^{2}\bar{\xi} \,] \eta = 0 \, ,
\end{equation}
%%%%
where $\bar{\xi}$ is the solution to (\ref{xi3:NC}). 

In order to study the NVE (\ref{etaNVE:NC}), we introduce the variable $z$ such that
%%%%
\begin{equation}
\label{varz:NC}
\cos\bar{\xi} = z \, .
\end{equation}
%%%%

With (\ref{varz:NC}) we may recast the NVE (\ref{etaNVE:NC}) as
%%%%
\begin{align}
\begin{split}
\eta''(z)+\qty(\frac{f'(z)}{2f(z)}+\frac{2}{z})\eta'(z) +\frac{\alpha_{2}}{f(z)}
\Big(\alpha_{2}z^{2}+\qty(\alpha_{4}-2\alpha_{6})\qty(1-z^{2})\Big)\eta(z)=0 \, ,
\end{split}\label{zNVE:NC}
\end{align}
%%%%
where
%%%%
\begin{equation}
\label{f:NC}
f(z) =\dot{\bar{\xi}}^{2}\sin^{2}\bar{\xi} = \qty(E+
\frac{\mathcal{A}_{\text{NC}}}{2}\qty(8z^{4}-8z^{2}
+1))\qty(1-z^{2}) \, ,
\end{equation}
%%%%
$E$ being the constant of integration equal to the energy of the string. We set $E=1$ in our analysis.

Next, we convert (\ref{zNVE:NC}) into the Schr\"{o}dinger form by using (\ref{Kov:Var}).
The resulting equation can be written as
%%%%
\begin{equation}
\label{formsch:NC}
\omega'(z)+\omega^{2}(z)=\frac{2B'(z)+B^{2}(z)-4A(z)}{4}
\equiv \mathcal{V}_{\text{NC}}(z) \, ,
\end{equation}
%%%%
where $\mathcal{V}_{\text{NC}}(z)$ is the Schr\"{o}dinger potential and
%%%%
\begin{subequations}
\label{AB:NC}
\begin{alignat}{2}
\begin{split}
A(z) &=\frac{\alpha_{2}}{f(z)}
\Big(\alpha_{2}z^{2}+\qty(\alpha_{4}-2\alpha_{6})\qty(1-z^{2})\Big) \, ,
\end{split}\label{A:NC}  \\[6pt]
\begin{split}
B(z) &= \qty(\frac{f'(z)}{2f(z)}+\frac{2}{z}) \, .
\end{split}\label{B:NC}
\end{alignat}
\end{subequations}
%%%%

The Schr\"{o}dinger potential $\mathcal{V}_{\text{NC}}(z)$ can be written as
%%%%
\begin{equation}
\label{formV:NC}
\mathcal{V}_{\text{NC}} = \frac{\mathcal{N}_{\text{NC}}}
{\mathcal{D}_{\text{NC}}} \, ,
\end{equation}
%%%%
where
%%%%
\begin{subequations}
\label{VND:NC}
\begin{alignat}{2}
\begin{split}
\mathcal{N}_{\text{NC}} &=12\qty(z^{2}-2)+\mathcal{A}^{2}_{\text{NC}}
\Big[ -54+536z^{2}+16z^{4}\qty(-147+263 z^{2}-208 z^{4}+60 z^{6}) \Big]
\\
&\qquad+ 4\mathcal{A}_{\text{NC}} \Big[ -30 +187 z^{2} -280 z^{4}+
120 z^{6}+2\alpha_{2} \qty(-1+9 z^{2}-16 z^{4}+8 z^{6})  \\
&\qquad\quad\quad\quad +16 \alpha_{2} \qty(z^{2}-1) \qty(z^{2}\alpha_{2}-
\qty(z^{2}-1)\alpha_{4}-2\alpha_{6}) \Big] \, ,
\end{split}\label{VN:NC}  \\[6pt]
\begin{split}
\mathcal{D}_{\text{NC}} &=4\qty(1-z^{2})^{2}\qty(2+\qty(1-8z^{2}
+8z^{4})\mathcal{A}_{\text{NC}})^{2}  \, .
\end{split}\label{VD:NC}
\end{alignat}
\end{subequations}
%%%%

In order to find the solution to (\ref{formsch:NC}), we now expand the potential
$\mathcal{V}_{\text{NC}}$ for small values of $z$ following the same argument that was
presented in the previous section \ref{ABJM:BD}. The resulting $\omega(z)$ equation may be
written as
%%%%
\begin{subequations}
\begin{alignat}{2}
\begin{split}
&~ \omega'(z)+\omega^{2}(z) \simeq  \widetilde{C}_{2}\, ,
\end{split}   \label{redSch:NC}   \\[5pt]  
\begin{split}
&~\widetilde{C}_{2} =~ -\frac{6+27\mathcal{A}_{\text{NC}}+4\alpha_{2} 
\qty(\alpha_{4}-2\alpha_{6})}{2\qty(2+\mathcal{A}_{\text{NC}})} \, .
\nonumber
\end{split}
\end{alignat}
\end{subequations}
%%%%
whose solution may be expressed as
%%%%
\begin{equation}
\label{solSch:NC}
\omega(z) = \sqrt{\widetilde{C}_{2}}~ \tanh \Big[ \sqrt{\widetilde{C}_{2}}
~\qty(z+\mathsf{C}_{2}) \Big] \, ,
\end{equation}
%%%%
where $\mathsf{C}_{2}$ is an arbitrary constant.

Now the Schr\"{o}dinger potential $\mathcal{V}_{\text{NC}}$ given by (\ref{formV:NC})
has poles of order $2$ at
%%%%
\begin{equation}\label{poles:NC}
z = \pm 1 \, , \quad z = \pm \frac{1}{2}\sqrt{2-\frac{\sqrt{2}
\sqrt{\mathcal{A}_{\text{NC}}\qty(\mathcal{A}_{\text{NC}}-2)}}
{\mathcal{A}_{\text{NC}}}} \, , \quad z = \pm \frac{1}{2}\sqrt{2+
\frac{\sqrt{2}\sqrt{\mathcal{A}_{\text{NC}}
\qty(\mathcal{A}_{\text{NC}}-2)}}{\mathcal{A}_{\text{NC}}}} \, .
\end{equation}
%%%%

On the other hand, the \emph{order at infinity} of $\mathcal{V}_{\text{NC}}$ is determined
to be $2$. These satisfy the criterion \textbf{Cd.}$(iii)$ of the Kovacic's algorithm as discussed
in Appendix \ref{Kova}. Also, for small $z$ the solution (\ref{solSch:NC}) is indeed a polynomial
of degree $1$. These information together ensure the analytic integrability of the system.

\subsubsection{Numerical results}\label{num:NC}
In order to numerically study the integrability of the string configuration, we use the ansatz 
(\ref{ansatz:NC}) together with the choice $\alpha_{i}=1$ of the winding numbers.

The resulting Hamilton's equations of motion are obtained as
%%%%
\begin{subequations}\label{Ham:NC}
\begin{alignat}{4}
\begin{split}
    \dot{\theta_{1}}=&\hspace{1mm}4p_{\theta_{1}}\sec^2\xi \, ,
\end{split} \label{Hamth:NC} \\[5pt]  
\begin{split}  
    \dot{\xi}=&\hspace{1mm}p_\xi  \, ,
\end{split}   \label{Hamxi:NC} \\[5pt]
\begin{split}    
\dot{p_{\theta_{1}}}=&\hspace{1mm}\frac{1}{2}\cos^2\xi\sin\theta_{1}
\qty(\sin^2\xi-\cos\theta_{1}\cos^2\xi) \, ,
\end{split}  \label{Hampth:NC}\\[5pt]
\begin{split}    
\dot{p_\xi}=&\hspace{1mm}\frac{1}{2}\qty(\cos\xi\sin^2\theta_{1}
\sin\xi - \cos^4\frac{\theta_{_1}}{2}\sin 4\xi
-8p^2_{\theta_{1}}\sec^2\xi\tan\xi )  \, ,
\end{split}  \label{Hampxi:NC}
\end{alignat}
\end{subequations}
%%%%
where we set $ \theta_2= p_{\theta_2}=0 $ in the rest of our analysis.

For numerical simulation, we set the following values of the Yang-Baxter deformation parameter: 
$\mu=0.01$, $ 0.8 $.

%%%%
\begin{figure}[t!]
\begin{multicols}{2}
    \includegraphics[width=\linewidth]{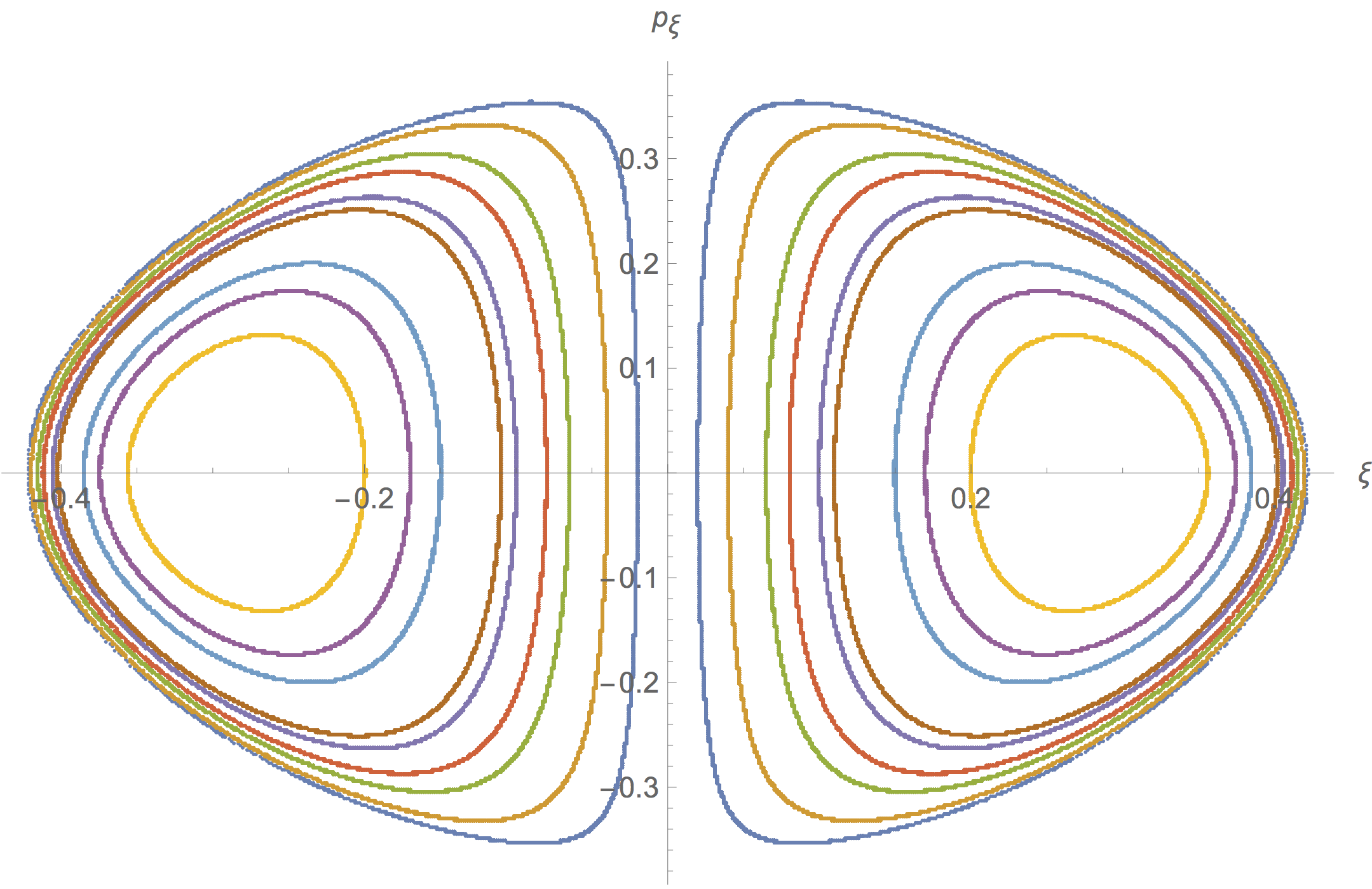}\par 
    \includegraphics[width=\linewidth]{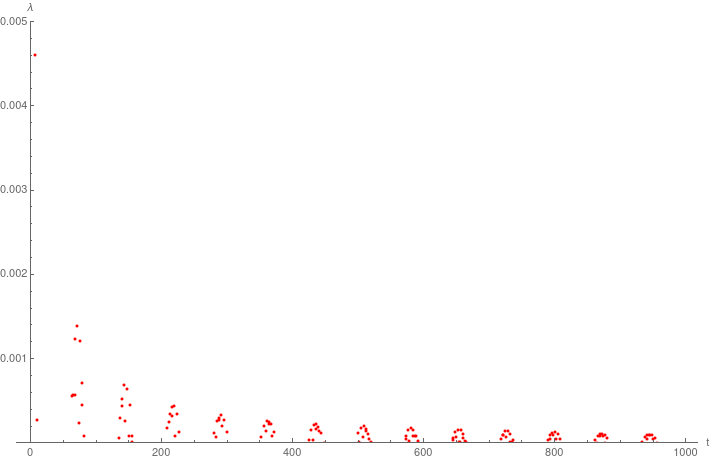}\par 
    \end{multicols}
\begin{multicols}{2}
    \includegraphics[width=\linewidth]{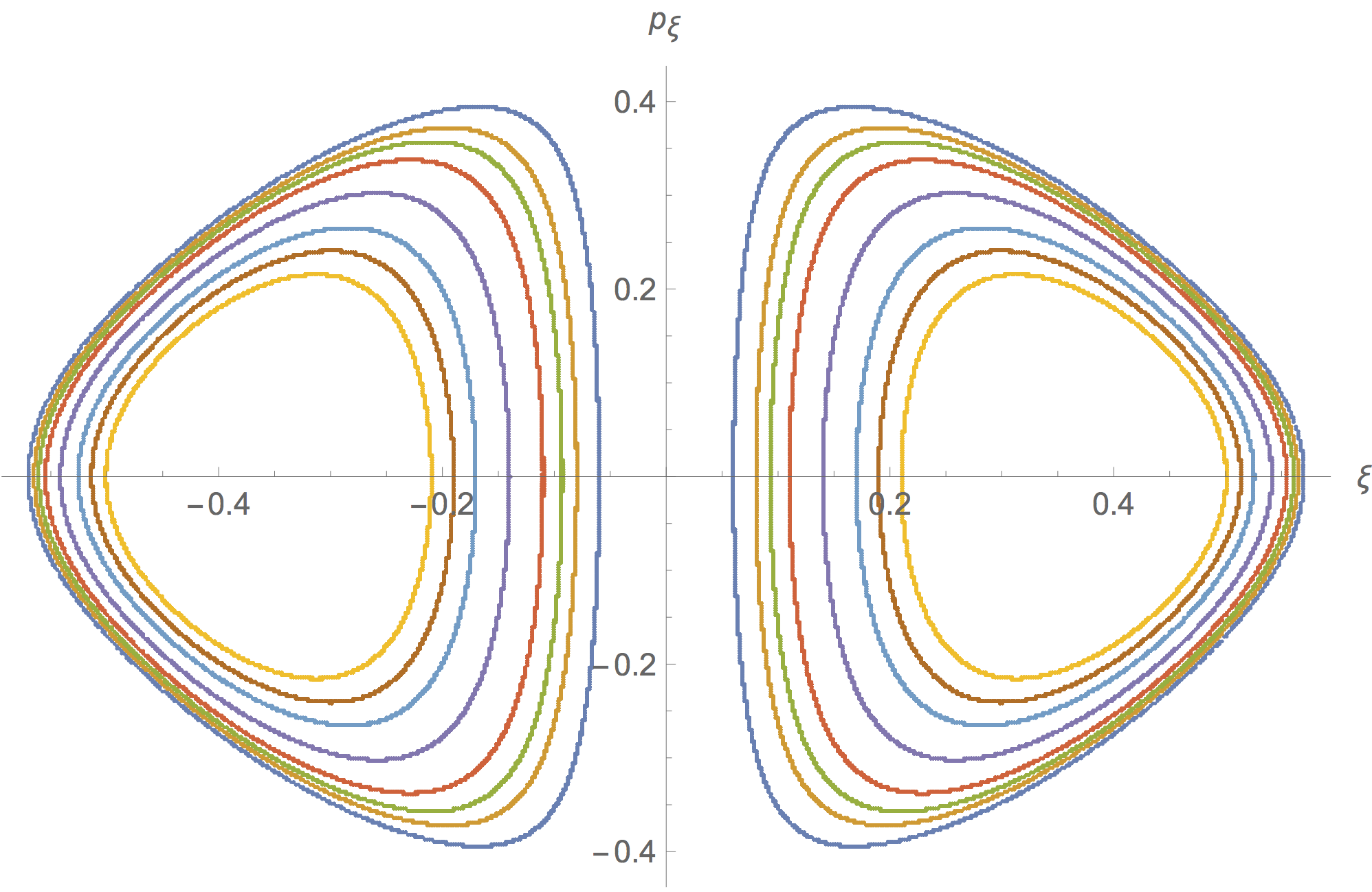}\par
    \includegraphics[width=\linewidth]{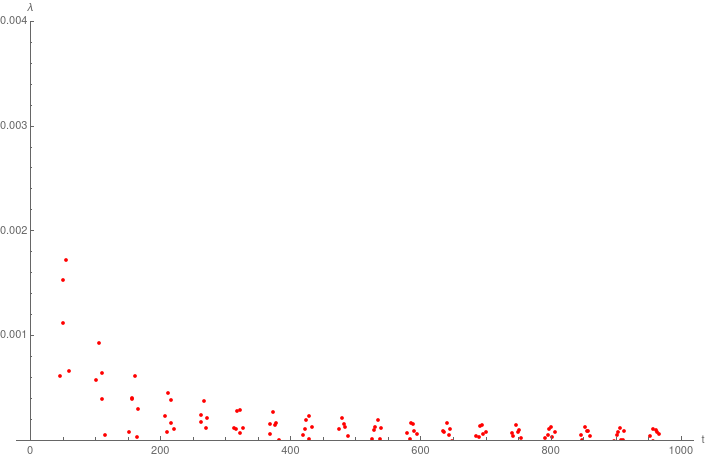}\par
\end{multicols}
\caption{Numerical plots of the Poincar\'{e} sections (\textbf{Left column}) and
Lyapunov exponents (\textbf{Right column}) for non-commutative ABJM. Here
we set the energy of the string $E_{0}=0.4$. The top plots are for
$\mu=0.01$ while the bottom plots are for $\mu=0.8$. The Poincar\'{e} sections
are nicely foliated KAM tori in the phase space and the Lyapunov exponent decays
to zero for large time $t$, indicating non-chaotic dynamics of the string.}
\label{fig2:NC}
\end{figure}
%%%%

Fig.\ref{fig2:NC} shows the corresponding Poincar\'{e} sections when the energy of the string is
$E=E_{0}=0.4$. Note that, we take the initial condition as $\theta_{1}(0)=0$, $p_\xi(0)=0$. We 
generate a random data set by choosing $\xi(0) \in [0, 1]$ which fixes the initial momenta 
$p_{\theta_{1}}(0)$ following the Hamiltonian constraints (\ref{Hamil}) and (\ref{app:Vir}). The
$\qty{ \xi,p_\xi }$ cross-section is obtained by collecting the data every time the trajectory
passes through the $ \theta_1=0 $ plane.

\begin{figure}[t]
	\begin{multicols}{2}
		\includegraphics[width=\linewidth]{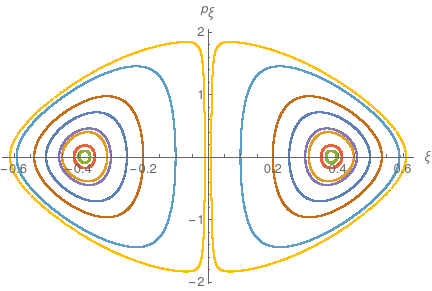}\par 
		\includegraphics[width=\linewidth]{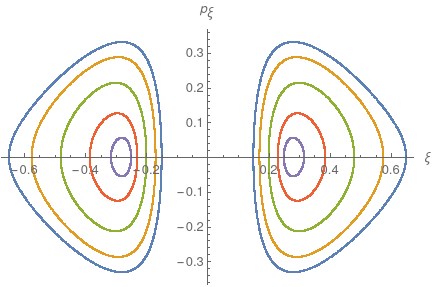}\par 
	\end{multicols}
	\caption{Additional plots of the Poincar\'{e} sections for non-commutative ABJM. On the left plot we set $\mathbf{E_{0}=1}$ and $ \mu=~0.5 $ whereas, the plot on the right corresponds to  $\mathbf{E_{0}=0.45}$, $ \mu =~0.1 $.}
	\label{Fig:New2:NC}
\end{figure}

In order to plot the Lyapunov exponent ($\lambda$) (fig.\ref{fig2:NC}), we set the initial conditions
as $\lbrace\theta_{1}(0)=0, \xi(0)=0.11, p_{\theta_{1}}(0)=0.17, p_\xi(0)=0 \rbrace $ together
with $\Delta X_{0}=10^{-7}$ in (\ref{app:Lyav}). For $\mu =0.8$ the initial conditions are changed
to $\qty{ \theta_{1}(0)=0, \xi(0)=0.11, p_{\theta_{1}}(0)=0.22, p_\xi(0)=0 } $. The energy of
these orbits are fixed at $ E=E_0 =0.4 $ such that, when put together, they satisfy the Hamiltonian 
constraints (\ref{Hamil}) and (\ref{app:Vir}). In the process, we finally generate a vanishing
Lyapunov exponent at large time ($t$) exhibiting a \emph{non-chaotic} motion. The validity of the above conclusions is further checked by plotting the Poincar\'{e} section for other values of the string energy, as shown in Fig.\ref{Fig:New2:NC}. 

%%%%
%%%%

%%%%%%%%%%%%%%%%%%%%%%%%%%%%%%%%%%
\subsection{Dipole deformed ABJM}\label{ABJM:DD}
Gravity dual of dipole deformed ABJM is obtained by considering a three parameter YB deformation
of AdS$_4$ $\times$ CP$^{3} $. The associated $ r $-matrix is constructed combining the
generators of both the AdS$_4$ and CP$^{3} $ subspaces\footnote{The $r$-matrix can be
written as
\begin{equation*}
r =~ \mathbf{p}_{2} \wedge \qty(\mu_{1} \mathbf{L}_{3}
+\mu_{2} \mathbf{L} + \mu_{3} \mathbf{M}_{3})  \, ,
\end{equation*}
where $\mathbf{L}=-1/\sqrt{3}\,\mathbf{L}_{8}+\sqrt{2/3}\,\mathbf{L}_{15}$ and
$\mathbf{L}_{3} \, , \mathbf{L}_{8} \, , \mathbf{L}_{15} \, , \mathbf{M}_{3}$ $\in$
$\mathfrak{su}(4)\oplus \mathfrak{su}(2)$ are Cartan generators. Here $\mu_{1}$,
$\mu_{2}$ and $\mu_{3}$ are deformation parameters in the theory \cite{Rado:2020swb}.
For this particular choice of the $r$-matrix, the deformation is along the $x_{2}$
direction in the AdS$_{4}$ and along the angular direction $\qty(\varphi_{1},\varphi_{2},
\psi)$ in CP$^{3}$.}.
The corresponding line element is given by \cite{Rado:2020swb}
%%%%
\begin{align}\label{met:DD}
\begin{split}
\dd s^{2}=&~ \frac{1}{4}\left(r^{2}\left(-\dd x_{0}^{2}+\dd x_{1}^{2}\right)
+\frac{r^{2}}{1+f_{3}^{2}} \dd x_{2}^{2}+\frac{\dd r^{2}}{r^{2}}\right)
+\dd \xi^{2}
\\
& +\frac{1}{4} \cos ^{2} \xi\left(\dd \theta_{1}^{2}+\sin ^{2} \theta_{1}
\dd \varphi_{1}^{2}\right)+\frac{1}{4} \sin ^{2} \xi\left(\dd \theta_{2}^{2}
+\sin ^{2} \theta_{2} \dd \varphi_{2}^{2}\right)
\\
&+\frac{1}{1+f_{3}^{2}}\left(\frac{1}{2} \cos\theta_{1} \dd \varphi_{1}
-\frac{1}{2} \cos \theta_{2} \dd \varphi_{2}+\dd \psi\right)^{2} \sin ^{2}\xi
\cos ^{2} \xi    \,  ,
\end{split}
\end{align}
%%%%
together with the NS-NS fluxes
%%%%
\begin{align}\label{ns:DD}
\begin{split}
B =& -\frac{1}{4}\left(\frac{f_{3}}{1+f_{3}^{2}}\right) r \dd x_{2}
\wedge\left(\frac{1}{2} \cos \theta_{1} \dd \varphi_{1}-\frac{1}{2}
\cos \theta_{2} \dd \varphi_{2}+\dd \psi\right) \sin \xi \cos \xi  \, ,
\\
f_{3} =& \frac{\mu r}{2} \sin (2\xi)   \, . 
\end{split} 
\end{align}
%%%%

We also set the Yang-Baxter deformation parameters as $\mu_{1}=\mu_{2}=0, \mu_{3}=\mu$.
Here, $\lbrace r, x_{0}, x_{1}, x_{2} \rbrace$ are the $AdS_4$ coordinates. On the other hand,
$\lbrace \xi, \theta_{1}, \theta_{2}, \varphi_{1}, \varphi_{2},\psi \rbrace$ are the coordinates
of internal $CP^3$ manifold. In the following analysis, we choose $x_0=t$,$x_1=$ constant and
$r=1$. 

In our analysis, we choose to work with the winding string ansatz of the form
%%%%
\begin{align}\label{ansatz:DD}
t =&t(\tau), \quad \theta_1=\theta_1(\tau),
\quad \theta_2=\theta_2(\tau),\quad \xi=\xi(\tau),\quad
\varphi_1=\alpha_2\sigma,\quad \nonumber\\[6pt]
\varphi_2 =&\beta_{2}\sigma, \quad \psi=\gamma_{2}
\sigma,\quad x_2=\eta_{2}\sigma  \, ,
\end{align}
%%%%
where $\alpha_{2}$, $\beta_{2}$, $\gamma_{2}$ and $\eta_{2}$ are the string winding
numbers.

Using the above ansatz (\ref{ansatz:DD}) we may write the Lagrangian in the action (\ref{POL})
as
%%%%
\begin{subequations}
\begin{alignat}{2}
\begin{split}
L_{P} &=~ -\frac{1}{2} \Bigg[ \frac{1}{4}\dot{t}^{2}-\dot{\xi}^{2}-\frac{1}{4}
		\dot{\theta}_{1}^{2}\cos^{2}\xi -\frac{1}{4}\dot{\theta}_{2}^{2}\sin^{2}\xi
		+\frac{x_2'^{2}}{4\qty(1+f_{3}^{2})} +\frac{\phi_{1}'^{2}\cos^{2}\xi}
		{4}
		\\
		&\quad \times \qty(\sin^{2}\theta_{1}+\frac{\sin^{2}\xi\cos^{2}\theta_{1}}
		{1+f_{3}^{2}})+\frac{\phi_{2}'^{2}\sin^{2}\xi}{4} 
		\qty(\sin^{2}\theta_{2}+\frac{\cos^{2}\xi\cos^{2}\theta_{2}}{1
			+f_{3}^{2}})
		\\
		&\quad  +\frac{\sin^{2}\xi \cos^{2}\xi}{1+f_{3}^{2}}
		\Big( \psi'^{2} -\frac{1}{2} \phi_{1}'\phi_{2}' \cos\theta_{1}
		\cos\theta_{2}+\phi_{1}'\psi'\cos\theta_{1}-\phi_{2}'\psi'
		\cos\theta_{2} \Big) \Bigg]
\end{split}     \label{act:DDa} \\[4pt]
\begin{split}
&= -\frac{1}{2} \Bigg[ \frac{1}{4}\dot{t}^{2}-\dot{\xi}^{2}-\frac{1}{4}
\dot{\theta}_{1}^{2}\cos^{2}\xi -\frac{1}{4}\dot{\theta}_{2}^{2}\sin^{2}\xi
+\frac{\eta_{2}^{2}}{4\qty(1+f_{3}^{2})} +\frac{\alpha_{2}^{2}\cos^{2}\xi}
{4}
\\
&\quad \times \qty(\sin^{2}\theta_{1}+\frac{\sin^{2}\xi\cos^{2}\theta_{1}}
{1+f_{3}^{2}})+\frac{\beta_{2}^{2}\sin^{2}\xi}{4} 
\qty(\sin^{2}\theta_{2}+\frac{\cos^{2}\xi\cos^{2}\theta_{2}}{1+f_{3}^{2}})
\\
&\quad  +\frac{\sin^{2}\xi \cos^{2}\xi}{1+f_{3}^{2}}
\Big( \gamma_{2}^{2} -\frac{1}{2} \alpha_{2}\beta_{2} \cos\theta_{1}
\cos\theta_{2}+\alpha_{2}\gamma_{2}\cos\theta_{1}-\beta_{2}\gamma_{2}
\cos\theta_{2} \Big) \Bigg]  \, .
\end{split}  \label{act:DD}
\end{alignat}
\end{subequations}
%%%%

\subsubsection{Analytical results}\label{ana:DD}
The eoms resulting from the variations of $\theta_{1}$, $\theta_{2}$ and $\xi$ in
(\ref{act:DD}) can be computed as\footnote{It is interesting to note that the order of the
YB deformation parameter that appear in the eoms is indeed $\order{\mu^{2}}$ and
no term of $\order{\mu}$ appears in the eoms. This is because the $B$ field does not
contribute to the eoms due to the choice of the ansatz (\ref{ansatz:DD}).}
%%%%
\begin{subequations}\label{eom:DD}
\begin{alignat}{2}
\begin{split}
0 =&~ 4\qty(1+\frac{\mu^{2}}{4} \sin^{2}2\xi)\qty(\cos\xi \, \ddot{\theta}_{1}
-2\sin\xi \, \dot{\xi} \, \dot{\theta}_{1})-4 \alpha_{2} \cos\xi \sin^{2}\xi
\qty(2 \gamma_{2}-\beta_{2}\cos\theta_{2})\sin\theta_{1} 
\\
&~ +2 \alpha_{2}^{2} \qty(1+\mu^{2} \sin^{2}\xi) \cos^{3}\xi \sin 2\theta_{1} \, ,
\end{split}  \label{t1:DD}  \\[6pt]
\begin{split}
0 =&~ 2\qty(1+\frac{\mu^{2}}{4} \sin^{2}2\xi)\qty(\sin\xi \, \ddot{\theta}_{2}
-2\cos\xi \, \dot{\xi} \, \dot{\theta}_{2}) + 2\beta_{2} \sin\xi \cos^{2}\xi
\big(2\gamma_{2}+\alpha_{2} \cos\theta_{1}
\\
&~ -\beta_{2} \cos\theta_{2} \big)\sin\theta_{2} + \beta_{2}^{2}
\qty(1+\frac{\mu^{2}}{4} \sin^{2}2\xi) \sin\xi \sin 2\theta_{2} \, ,
\end{split}  \label{t2:DD}  \\[6pt]
\begin{split}
0 =&~ \qty(1+\frac{\mu^{2}}{4} \sin^{2}2\xi)^{2} \qty(16 \ddot{\xi}
+2 \sin 2\xi \qty(\dot{\theta}_{1}^{2}-\dot{\theta}_{2}^{2}- \beta_{2}^{2}
\sin^{2}\theta_{2} - \alpha_{2}^{2} \sin^{2}\theta_{1}))
\\
&~ +\frac{\sin 4\xi}{2} \Big( 4\gamma_{2}^{2} - \mu^{2}\eta_{2}^{2}
+\alpha_{2}^{2} \cos^{2}\theta_{1} + \beta_{2}^{2} \cos^{2}\theta_{2}
-4 \beta_{2}\gamma_{2} \cos\theta_{2}
\\
&~ + 2\alpha_{2} \cos\theta_{1}\qty(2 \gamma_{2} - \beta_{2} \cos\theta_{2})
\Big)  \, .
\end{split}  \label{xi:DD} 
\end{alignat}
\end{subequations}
%%%%

We observe that the conjugate momenta corresponding to the coordinates $\{t, \Phi_{i}\}$
with $(i=\varphi_{1},\varphi_{2},\psi,x_{2})$ can be computed as
%%%%
\begin{equation}\label{conmom:DD}
E \equiv \frac{\partial L_{P}}{\partial\dot{t}} =-\frac{\dot{t}}{4} \, ,
\qquad P_{\Phi_{i}} \equiv \frac{\partial L_{P}}{\partial\dot{\Phi_{i}}}
=0 \, ,
\end{equation} 
%%%%
which are indeed found to be conserved
%%%%
\begin{equation}\label{ConChg:DD}
\partial_{\tau}E=0 \, \quad (\text{in $t=\tau$ gauge}) \, ,
\qquad \partial_{\tau} P_{\Phi_{i}}=0  \, .
\end{equation}
%%%%

Now using the definition (\ref{EM:gen}) of the energy-momentum tensor $T_{ab}$, it is easy to
check that
%%%%
\begin{alignat}{2}
\begin{split}
\partial_{\tau}T_{\tau\tau} &= 0 \, ,  \quad \text{on-shell} \, ,
\\[4pt]
\partial_{\tau}T_{\tau\sigma} &= 0 \, ,  \quad \text{trivially} \, .
\end{split}   \label{EM:DD}
\end{alignat}
%%%%

In the next step, we study the dynamics of the string governed by the eoms (\ref{eom:DD}). In
our analysis we first choose the $\theta_{2}$ invariant plane in the phase space defined as
%%%%
\begin{equation}
\label{IP1:DD}
\theta_{2}\sim 0 \, \qquad \Pi_{\theta_{2}}:=\dot{\theta}_{2}\sim 0 \, .
\end{equation}
%%%%
This choice trivially satisfies the $\theta_{2}$ eom (\ref{t2:DD}), and the remaining two eoms
(\ref{t1:DD}), (\ref{xi:DD}) reduce to
%%%%
\begin{subequations}
\begin{alignat}{2}
\begin{split}
0 =&~ 4\qty(1+\frac{\mu^{2}}{4} \sin^{2}2\xi)\qty(\cos\xi \, \ddot{\theta}_{1}
-2\sin\xi \, \dot{\xi} \, \dot{\theta}_{1})-4 \alpha_{2} \cos\xi \sin^{2}\xi
\qty(2 \gamma_{2}-\beta_{2})\sin\theta_{1} 
\\
&~ +2 \alpha_{2}^{2} \qty(1+\mu^{2} \sin^{2}\xi) \cos^{3}\xi \sin 2\theta_{1} \, ,
\end{split}  \label{t12:DD} \\[6pt]
\begin{split}
0 =&~ \qty(1+\frac{\mu^{2}}{4} \sin^{2}2\xi)^{2} \qty(16 \ddot{\xi}
+2 \sin 2\xi \qty(\dot{\theta}_{1}^{2} - \alpha_{2}^{2}\sin^{2}\theta_{1}))
\\
&~ +\frac{\sin 4\xi}{2} \Big( 4\gamma_{2}^{2} - \mu^{2}\eta_{2}^{2}
+\alpha_{2}^{2} \cos^{2}\theta_{1} + \beta_{2}^{2} -4 \beta_{2}\gamma_{2}
+ 2\alpha_{2} \cos\theta_{1}\qty(2 \gamma_{2} - \beta_{2})
\Big)  \, . 
\end{split}  \label{xi2:DD} 
\end{alignat}
\end{subequations}
%%%%

The dynamics of the string in the reduced phase space, governed by (\ref{t12:DD}) and
(\ref{xi2:DD}), can be studied by further choosing the $\theta_{1}$ invariant plane defined
as
%%%%
\begin{equation}\label{IP2:DD}
\theta_{1} \sim 0 \, , \qquad  \Pi_{\theta_{1}}
\equiv \dot{\theta}_{1} \sim 0 \, .
\end{equation}
%%%%
While is choice trivially satisfies (\ref{t12:DD}), (\ref{xi2:DD}) reduces to the form
%%%%
\begin{equation}\label{xi3:DD}
16\qty(1+\frac{\mu^{2}}{4}\sin^{2}2\xi)^{2} \ddot{\xi}
+\qty( \mathcal{A}_{\text{DD}} - \mu^{2} \eta_{2}^{2})
\sin 4\xi = 0  \, ,
\end{equation}
%%%%
where
%%%%
\begin{equation}\label{ANC:DD}
\mathcal{A}_{\text{DD}} = \frac{1}{2} \Big( 4\gamma_{2}^{2}
+\alpha_{2}^{2} + \beta_{2}^{2} -4\beta_{2}\gamma_{2}+
2\alpha_{2}\qty(2 \gamma_{2} - \beta_{2}) \Big) \, .
\end{equation}
%%%%

Next we consider infinitesimal fluctuation ($\delta\theta_{1} \sim \eta$) around the $\theta_{1}$
invariant plane. This results in the normal variational equation (NVE) equation which can be
written as
%%%%
\begin{align} \label{etaNVE:DD}
\begin{split}
& \qty(1+\frac{\mu^{2}}{4} \sin^{2}2\xi)\qty(\cos\xi \, \ddot{\eta}
-2\sin\xi \, \dot{\xi} \, \dot{\eta}_{1})
\\
& \qquad\qquad +\Big( \alpha_{2}^{2} \qty(1+\mu^{2} \sin^{2}
\bar{\xi})\cos^{3}\bar{\xi} - \alpha_{2} \qty(2 \gamma_{2}
-\beta_{2})\cos\bar{\xi} \sin^{2}\bar{\xi}   \Big) \eta =~ 0  \, .
\end{split}
\end{align} 
%%%%

Using the change in variable $\displaystyle \cos\bar{\xi} = z$ we may recast (\ref{etaNVE:DD})
in the form
%%%%
\begin{equation}\label{zNVE:DD}
\eta''(z)+B(z)\eta'(z)+A(z) = 0  \, ,
\end{equation}
%%%%
where
%%%%
\begin{subequations}
\begin{alignat}{2}
\begin{split}
B(z) =&~ \frac{f'(z)}{2f(z)}+\frac{2}{z}  \, ,
\end{split}  \label{B:DD} \\[5pt]
\begin{split}
A(z) =&~ \frac{\alpha_{2}^{2}z^{2}\qty(1+\mu^{2}\qty(1-z^{2}))
-\alpha_{2}\qty(2\gamma_{2}-\beta_{2})\qty(1-z^{2})}
{\Big(1+\mu^{2}z^{2}\qty(1-z^{2})\Big) f(z)}  \, ,
\end{split}  \label{A:DD} \\[5pt]
\begin{split}
f(z) =&~  \dot{\bar{\xi}}^{2} \sin^{2}\bar{\xi}
\\
=&~ \Bigg( E+\frac{\mathcal{A}_{\text{DD}}-\mu^{2}\eta_{2}^{2}}{32}
\qty(8z^{4}-8z^{2}+1) - \frac{\mu^{2}\mathcal{A}_{\text{DD}}}{128}
\qty(-128 z^{8}+256 z^{6}-152 z^{4} +24 z^{2}) \Bigg)
\\
&~ \times \qty(1-z^{2}) \, .
\end{split}  \label{f:DD}
\end{alignat}
\end{subequations}
%%%%

In (\ref{f:DD}) $E$ is the energy of the propagating string and we choose $E=1$ without any loss
of generality. Also notice that, in deriving (\ref{f:DD}) we series expand (\ref{xi3:DD}) for small
values of the YB parameter $\mu$ and keep terms upto $\order{\mu^{2}}$.

We can further recast (\ref{zNVE:DD}) in the Schr\"{o}dinger form (\ref{Kov:Fnl}) using
(\ref{Kov:Var}). The resulting equation can then be written as
%%%%
\begin{equation}
\label{formsch:DD}
\omega'(z)+\omega^{2}(z)=\frac{2B'(z)+B^{2}(z)-4A(z)}{4}
\equiv \mathcal{V}_{\text{DD}}(z) \, ,
\end{equation}
%%%%
where $\mathcal{V}_{\text{DD}}(z)$ is the Schr\"{o}dinger potential whose exact form is
quite complicated and we avoid writing the detailed expression here. However, we can expand
this potential $\mathcal{V}_{\text{DD}}(z)$ for small $\mu$ as well as small $z$. The latter
expansion is justified whenever we work with the full CP$^{3}$ metric (\ref{met:DD}). The
final form of (\ref{formsch:DD}) can be computed as
%%%%
\begin{equation}
\label{redSch:DD}
\omega'(z)+\omega^{2}(z) \approx \widetilde{C}_{3} \, ,
\end{equation}
%%%%
where 
%%%%
\begin{align}
\begin{split}
\widetilde{C}_{3} =&~ - \frac{96+27 \mathcal{A}_{\text{DD}}
+64\alpha_{2}\beta_{2}-128\alpha_{2}\gamma_{2}}{2
\qty(32+\mathcal{A}_{\text{DD}})}
\\
&~ + \frac{\qty(-288\mathcal{A}_{\text{DD}} -9 \mathcal{A}_{\text{DD}}^{2}
+384 \eta_{2}^{2}-32\alpha_{2}\beta_{2}\eta_{2}^{2}+64\alpha_{2}
\gamma_{2}\eta_{2}^{2} )\mu^{2}}{\qty(32+\mathcal{A}_{\text{DD}})^{2}}
+\order{\mu^{4}}  \, .
\end{split}
\end{align}
%%%%

The general solution to (\ref{redSch:DD}) may be obtained as
%%%%
\begin{equation}\label{solSch:DD}
\omega(z) = \sqrt{\widetilde{C}_{3}}~ \tanh \qty(\sqrt{\widetilde{C}_{3}}
~\qty(z+\mathsf{C}_{3}))   \, ,
\end{equation}
%%%%
where $\mathsf{C}_{3}$ is the integration constant. Note that, for small $z$ the solution
(\ref{solSch:DD}) is indeed a polynomial of degree $1$. Moreover, there are poles of order
$2$ of the potential $\mathcal{V}_{\text{DD}}(z)$ at\footnote{The detailed expressions
of the poles at $z_{i}$ are not important in our discussion. Hence we avoid writing their
forms here.}
%%%%
\begin{equation}
\label{poles:DD}
z=\pm 1 \, , \qquad z = z_{i} \, \quad i=1,\cdots,4  \, ,
\end{equation}
%%%%
and the order at infinity of $\mathcal{V}_{\text{DD}}(z)$ is $2$. Thus the criterion
\textbf{Cd}(iii) of the Kovacic's algorithm, discussed in Appendix \ref{Kova}, is satisfied.
From these results we can infer that the system is indeed integrable.

\subsubsection{Numerical results}\label{num:DD}
We now explore the non-chaotic dynamics of the string configuration using numerical methods.
In order to do so, we note down the corresponding Hamilton's equations of motion\footnote{We
choose, $ \theta_2= p_{\theta_2}=0 $ as before.}
%%%%
\begin{subequations}
\begin{alignat}{2}
\begin{split}
    \dot{\theta_{1}}=&\hspace{1mm}4p_{\theta_{1}}\sec^2\xi  \, ,
\end{split}  \label{Hamilth:DD}     \\[4pt]
\begin{split}
    \dot{\xi}=&\hspace{1mm}p_\xi  \, ,
\end{split}  \label{Hamilxi:DD}     \\[4pt] 
\begin{split}   
     \dot{p_{\theta_{1}}}= &\hspace{1mm}\frac{  \cos^4\xi(-2-\mu^2+\mu^2\cos(2\xi))\sin(2\theta_{1})+\sin\theta_{1}\sin^2(2\xi)}{8+2\mu^2\sin^2(2\xi)}   \, ,
\end{split}  \label{Hamilpth:DD}     \\[4pt]      
\begin{split}   
     \dot{p_\xi}=&\hspace{1mm}\frac{\mathcal{N}_2}{8(4+\mu^2\sin^2(2\xi))^2}  \, ,
\end{split}  \label{Hamilpxi:DD}     
\end{alignat}
\end{subequations}
%%%%
where we denote
    \begin{align}
  \mathcal{N}_2=&\hspace{1mm}\bigg(-32\big(3-2\mu^2+4\cos\theta_{1}
  +\cos{(2\theta_{1})}\big)\cos(2\xi)-8\big(8+\mu^2-\mu^2\cos(4\xi)
  \big)^2p^2_{\theta_{1}}\sec^4\xi\nonumber\\
  &+\big(8+\mu^2-\mu^2\cos(4\xi)\big)^2\sin^2\theta_{1}\bigg)\cos\xi\sin\xi.
\end{align}
%%%%

%%%%
\begin{figure}[t]
\begin{multicols}{2}
    \includegraphics[width=\linewidth]{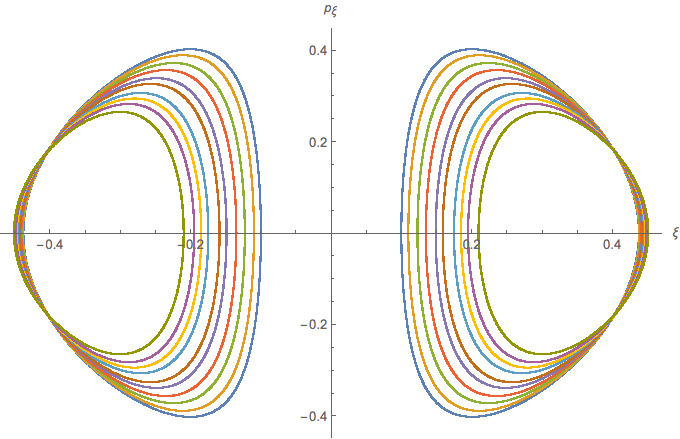}\par 
    \includegraphics[width=\linewidth]{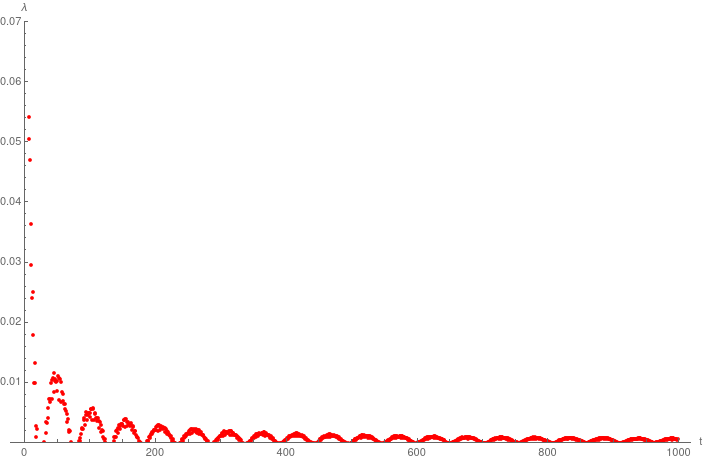}\par 
    \end{multicols}
\begin{multicols}{2}
    \includegraphics[width=\linewidth]{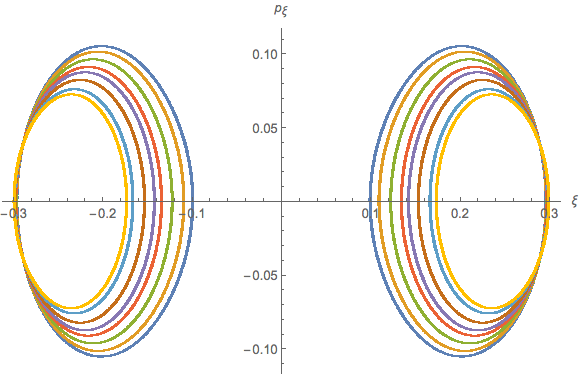}\par
    \includegraphics[width=\linewidth]{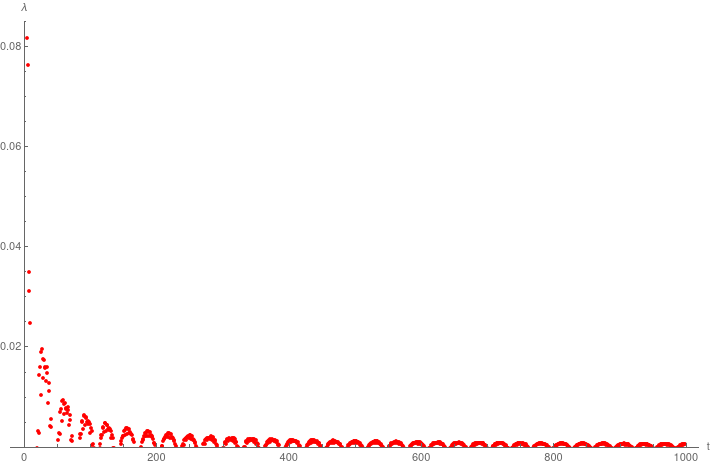}\par
\end{multicols}
\caption{Numerical plots of the Poincar\'{e} sections (\textbf{Left column}) and
Lyapunov exponents (\textbf{Right column}) for dipole deformed ABJM. Here
we set the energy of the string $E_{0}=0.35$. The top plots are for
$\mu=0.01$ while the bottom plots are for $\mu=0.8$. The Poincar\'{e} sections
can be seen to be undistorted foliations of KAM tori in the phase space and for
large time $t$ the Lyapunov exponent decays to zero. These are indications of the
non-chaotic dynamics of the string configuration.}
\label{fig3:DD}
\end{figure}
%%%%

In order to obtain the Poincar\'{e} sections, we set the energy as $ E=E_0=0.35 $ while the rest
of the data is chosen as $\theta_{1}(0)=0$ and $p_\xi(0)=0$. Given this initial data, we generate
a random data set for $p_{\theta_{1}}(0)$ by choosing $\xi (0)\in [0, 1]$ such that the constraints 
(\ref{app:Vir}) are satisfied. We also set the values of the winding numbers in (\ref{ansatz:DD}) as
$\alpha_{2}=\beta_{2}=\gamma_{2}=\eta_{2}=1$.

In our numerical analysis, the YB parameter is set to be, $ \mu =0.01 $ and $ 0.8 $. As in the
previous cases, the Poincar\'{e} sections (Fig.\ref{fig3:DD}) are obtained by plotting all the points
those are on the $\lbrace\xi, p_\xi\rbrace$ plane which correspond to trajectories passing through 
the $\theta_{1}=0$ hyper-plane. 

In order to calculate the Lyapunov exponent ($\lambda$), we set the initial conditions as
$\theta_{1}(0)=0$, $p_\xi(0)=0$, $\xi (0)=0.1$ and $p_{\theta_{1}}(0)=0.23$ those are
compatible with the Hamiltonian constraint (\ref{app:Vir}). The initial separation between the two
nearby trajectories is set to be $ \Delta X_{0}=10^{-7} $ as before, which eventually results in a
zero value for the Lyapunov (Fig.\ref{fig3:DD}) for large $t$. For YB parameter value $\mu =  0.8 $,
the initial data are set to be $\theta_{1}(0)=0$, $p_\xi(0)=0$, $\xi (0)=0.2$ and $p_{\theta_{1}}
(0)=0.22$.

\begin{figure}[t]
	\begin{multicols}{2}
		\includegraphics[width=\linewidth]{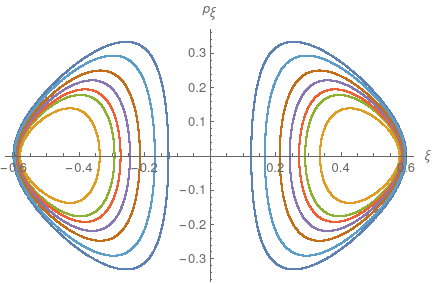}\par 
		\includegraphics[width=\linewidth]{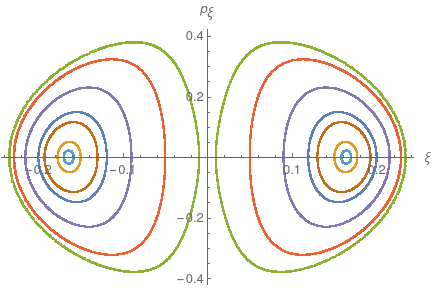}\par 
	\end{multicols}
	\caption{Additional plots of the Poincar\'{e} sections for dipole deformed ABJM. On the left plot we set $\mathbf{E_{0}=1}$ and $ \mu=~0.5 $, and on the right plot we set  $\mathbf{E_{0}=0.55}$, $ \mu =~0.1 $.}
	\label{Fig:New3:DD}
\end{figure}

Clearly, the nicely foliated KAM tori trajectories in the phase space along with the vanishing Lyapunov exponent indicate non-chaotic dynamics of the superstring propagating in this deformed background. We further plot Poincar\'{e} sections corresponding to two different energies ($ E=0.55 $ and $ E=1 $) of the string in Fig.\ref{Fig:New3:DD}. In these cases we observe that the KAM tori trajectories are nicely foliated as well, ruling out the chaotic behaviour of the string configuration.

%%%%
%%%%

%%%%%%%%%%%%%%%%%%%%%%%%%%%%%%%%%%
\subsection{Nonrelativistic ABJM}\label{ABJM:NR}
The gravity dual of nonrelativistic ABJM is obtained by constructing Abelian $ r $-matrices using
Cartan generators of both AdS$_{4}$ as well as CP$^{3}$ subspaces\footnote{The
$ r $-matrix is written as
\begin{equation*}
r=~ \mathbf{p}_{-} \wedge \qty(\mu_{1} \mathbf{L}_{3}
+\mu_{2} \mathbf{L} + \mu_{3} \mathbf{M}_{3})  \, ,
\end{equation*}
where $\mu_{i}$ are the YB deformation parameters and $\mathbf{p}_{\pm}=~ 
\qty(\mathbf{p}_{0}\pm \mathbf{p}_{2})\big/ \sqrt{2}$ are the light-cone momenta
corresponding to the light-cone coordinates (\ref{LC:NR}) \cite{Rado:2020swb}.}. The
corresponding line element is given by \cite{Rado:2020swb}
%%%%
\begin{align}\label{met:NR}
\begin{split}
\dd s^{2}=& \frac{1}{4}\left(-2 r^{2} d x_{+} d x_{-}+r^{2} d x_{1}^{2}
+\frac{d r^{2}}{r^{2}}-\mathcal{M} r^{2} d x_{+}^{2}\right)
+\dd s_{C P^{3}}^{2}  \, ,    \\[5pt]
\mathrm{d} s^{2}_{CP^3}=&\hspace{1mm}\mathrm{d}\xi^2+\frac{1}{4}\cos^2{\xi}
(\mathrm{d}\theta_1^2+\sin^2{\theta_1}\mathrm{d}\phi_1^2)+\frac{1}{4}\sin^2{\xi}
(\mathrm{d}\theta_2^2+\sin^2{\theta_2}\mathrm{d}\phi_2^2)    \\
&\quad  +\bigg(\frac{1}{2}\cos{\theta_1}\mathrm{d}\phi_1-\frac{1}
{2}\cos{\theta_2}\mathrm{d}\phi_2+\mathrm{d}\psi\bigg)^2\sin^2{\xi}\cos^2{\xi}  \, ,
\end{split}
\end{align}
%%%%
where 
%%%%
\begin{align}
\begin{split}
\mathcal{M} &=f_{1}^{2}+f_{2}^{2}+f_{3}^{2}  \, , \\
f_{1} &=\frac{r}{2 \sqrt{2}} \mu_{1} \sin \theta_{1} \cos \xi   \, ,\\
f_{2} &=\frac{r}{2 \sqrt{2}} \mu_{2} \sin \theta_{2} \sin \xi   \, , \\
f_{3} &=\frac{r}{2 \sqrt{2}}\left(2 \mu_{3}-\mu_{1} \cos \theta_{1}
+\mu_{2} \cos \theta_{2}\right) \sin \xi \cos \xi  \, .
\end{split}
\end{align}
%%%%

The corresponding $B$-field may be written as
%%%%
\begin{align}\label{ns:NR}
\begin{split}
B=&-\frac{1}{\sqrt{2}} r \cos \xi\left(f_{1} \sin \theta_{1}-f_{3}
\cos \theta_{1} \sin \xi\right) d x_{+} \wedge d \varphi_{1} \\
&-\frac{1}{\sqrt{2}} r \sin \xi\left(f_{3} \cos \theta_{2} \cos \xi
+f_{2} \sin \theta_{2}\right) d x_{+} \wedge d \varphi_{2} \\
&+\frac{1}{\sqrt{2}} r \sin (2 \xi) f_{3} d x_{+} \wedge d \psi  \, .
\end{split}
\end{align}
%%%%

The light-cone coordinates $x_{\pm}$ appearing in (\ref{met:NR}) and (\ref{ns:NR}) are
given by
%%%%
\begin{align}\label{LC:NR}
 x_{\pm}=\frac{1}{\sqrt{2}}\left(x^{0} \pm x^{2}\right)  \, .
\end{align}
%%%%

Notice that, in (\ref{met:NR}) $\mu_{i}$ $(i=1,2,3)$ are the Yang-Baxter (YB) deformation
parameters of the theory. Also, the metric (\ref{met:NR}) corresponds to a Schr\"{o}dinger
space-time with dynamical critical exponent $2$. In the following analytical and numerical
analyses, we choose the AdS$_{4}$ coordinates as $x_0=t$, $r=1$ and $x_1=$constant.

We now work with the winding string ansatz of the form
%%%%
\begin{align}\label{ansatz:NR}
    x_{+} &= ~ x_{+}(\tau) \, , & \theta_{1} &=\theta_1(\tau) \, ,
    & \theta_{2} &=\theta_{2}(\tau) \, ,  & \xi &=\xi(\tau)\, ,
    \\\nonumber
    \phi_{1} &=\alpha_{2}\sigma  \, ,
    &    \phi_{2} &=\alpha_{4}\sigma \, ,
    &    \psi &= \alpha_{6}\sigma  \, ,
    &    x_{-} &= \eta_{1}\tau   \, .
\end{align}
%%%%

Here $\alpha_{2}$, $\alpha_{4}$, $\alpha_{6}$ and $\eta_{1}$ are the winding numbers of
the string.

Using (\ref{ansatz:NR}), the Lagrangian in the Polyakov action (\ref{POL}) can be written as
%%%%
\begin{subequations}
\begin{alignat}{2}
\begin{split}
	L_{P} &=~ -\frac{1}{2}\Bigg[ \frac{\mathcal{M}}{4}\dot{x}_{+}^{2}
		-\dot{\xi}^{2} -\frac{1}{4}\qty(\dot{\theta}_{1}^{2}\cos^{2}\xi +
		\dot{\theta}_{2}^{2}\sin^{2}\xi)  +\frac{\phi_{1}'^2}{4}\cos^{2}\xi 
		\qty(\sin^{2}\theta_{1}+\sin^{2}\xi \cos^{2}\theta_{1})
		\\
		& \quad  +\frac{\phi_{2}'^2}{4}\sin^{2}
		\xi \qty(\sin^{2}\theta_{2}+\cos^{2}\xi \cos^{2}\theta_{2}) +\sin^{2}\xi
		\cos^{2}\xi \Big(\psi'^2+\phi_{1}'
		\psi'\cos\theta_{1}-\phi_{2}'
		\psi'\cos\theta_{2}
		\\	
		& \quad -\frac{1}{2} \phi_{1}'
		\phi_{2}' \cos\theta_{1} \cos\theta_{2} \Big)
		\Bigg] 
		-\frac{\phi_{1}'}{\sqrt{2}}\dot{x}_{+} \cos\xi \qty(f_{1}\sin\theta_{1}-f_{3}\sin\xi
		\cos\theta_{1})
		\\
		& \quad - \frac{\phi_{2}'}{\sqrt{2}}\dot{x}_{+} \sin\xi \qty(f_{2}\sin\theta_{2}
		+f_{3}\cos\xi	\cos\theta_{2})\dot{x}_{+} +\frac{\psi'}{\sqrt{2}} \dot{x}_{+}
		 f_{3}\sin2\xi 
	\end{split}    \label{act:NRa}  \\[6pt]
	\begin{split}
	&=~ -\frac{1}{2}\Bigg[ \frac{\mathcal{M}}{4}\dot{x}_{+}^{2}
	-\dot{\xi}^{2} -\frac{1}{4}\qty(\dot{\theta}_{1}^{2}\cos^{2}\xi +
	\dot{\theta}_{2}^{2}\sin^{2}\xi)  +\frac{\alpha_{2}^{2}}{4}\cos^{2}\xi 
	\qty(\sin^{2}\theta_{1}+\sin^{2}\xi \cos^{2}\theta_{1})
	\\
	& \quad  +\frac{\alpha_{4}^{2}}{4}\sin^{2}
	\xi \qty(\sin^{2}\theta_{2}+\cos^{2}\xi \cos^{2}\theta_{2}) +\sin^{2}\xi
	\cos^{2}\xi \Big(\alpha_{6}^{2}+\alpha_{2}
	\alpha_{6}\cos\theta_{1}-\alpha_{4}\alpha_{6}\cos\theta_{2}
	\\	
	& \quad -\frac{1}{2} \alpha_{2}\alpha_{4} \cos\theta_{1} \cos\theta_{2} \Big)
	\Bigg] 
	-\frac{\alpha_{2}}{\sqrt{2}}\dot{x}_{+} \cos\xi \qty(f_{1}\sin\theta_{1}-f_{3}\sin\xi
	\cos\theta_{1})
	\\
	& \quad - \frac{\alpha_{4}}{\sqrt{2}} \dot{x}_{+}\sin\xi \qty(f_{2}\sin\theta_{2}
	+f_{3}\cos\xi \cos\theta_{2}) +\frac{\alpha_{6}}{\sqrt{2}}\dot{x}_{+} f_{3}\sin2\xi  \, .
\end{split}  \label{act:NR}
\end{alignat}
\end{subequations}
%%%%
Notice that, even with the ansatz (\ref{ansatz:NR}), there exists a non-trivial contribution of the
$B$-field in the dynamics of the string unlike the previous cases in Sections \ref{ABJM:BD}, 
\ref{ABJM:NC} and \ref{ABJM:DD}.

\subsubsection{Analytical results}\label{ana:NR}
We first use (\ref{act:NR}) to compute the eom corresponding to the coordinate $x_{+}$. The
result can be written as
%%%%
\begin{equation}\label{eom:xp}
\mathcal{M}\frac{\dd}{\dd \tau}\dot{x}_{+}+\dot{x}_{+}
\frac{\dd}{\dd \tau}\big( \mathcal{M} - 4 B_{+j} \big) =~0 \, ,
\end{equation}
%%%%
where $B_{+j}$ ($+\equiv x_{+}$, $j=\varphi_{1},\varphi_{2},\psi$) are the components of
the $B$ field in (\ref{ns:NR}) which also appear in the Lagrangian (\ref{act:NR}).

We can get rid of the first term in the LHS of (\ref{eom:xp}) by choosing
%%%%
\begin{equation}\label{def:xp}
x_{+}=\mathscr{E}\tau  \, .
\end{equation}
%%%%
We can always set $\mathscr{E}=1$ without loss of any generality. Eq.(\ref{def:xp}) shows
that $x_{+}$ is indeed the world-sheet time. However, the vanishing of the second term in
the LHS of (\ref{eom:xp}) results in the following constraint equation:
%%%%
\begin{align}\label{tCon:NR}
\begin{split}
& \dot{\xi} \Big[ \qty(\mu_{2}^{2}\sin^{2}\theta_{2} -\mu_{1}^{2}\sin^{2}
\theta_{1})\sin 2\xi + \qty(2\mu_{3}-\mu_{1}\cos\theta_{1}+\mu_{2}
\cos\theta_{2})^{2}\cos 2\xi \Big]
\\
& + \dot{\theta}_{1} \Big[ \mu_{1}^{2} \sin 2\theta_{1}
\cos^{2}\xi +\mu_{1} \sin\theta_{1}\qty(2\mu_{3}-\mu_{1}\cos\theta_{1}
+\mu_{2}\cos\theta_{2}) \sin 2\xi \Big] 
\\
&+ \dot{\theta}_{2} \Big[ \mu_{2}^{2}
\sin 2\theta_{2}\sin^{2}\xi -\mu_{2} \sin\theta_{2}\qty(2\mu_{3}-\mu_{1}
\cos\theta_{1}+\mu_{2}\cos\theta_{2}) \sin 2\xi \Big]   =~0  \, .
\end{split}
\end{align}
%%%%

Next we again use (\ref{act:NR}) to derive the eoms corresponding to the $\theta_{1}$,
$\theta_{2}$ and $\xi$ coordinates. The results are formally written as
%%%%
\begin{subequations}\label{eom:NR}
\begin{alignat}{3}
\begin{split}
0 =&~ 8 \cos\xi ~ \ddot{\theta}_{1}- 16 \sin\xi ~\dot{\xi}\dot{\theta}_{1}
-\sin^{2}\xi \cos\xi \sin\theta_{1} \Big(16 \mu_{1} \alpha_{6}-8 \mu_{1}
\alpha_{2} \cos\theta_{1} + 8\mu_{1} \alpha_{4} \cos\theta_{2}
\\
&~ -\qty(\mu_{1}+8\alpha_{2})\qty(-2\mu_{3}+\mu_{1} \cos\theta_{1}
-\mu_{2} \cos\theta_{2})+8\qty(-2 \alpha_{2}\alpha_{6}+\alpha_{2}
\alpha_{4} \cos\theta_{2}) \Big) 
\\
&~ + \cos\xi \sin 2\theta_{1} \qty(8 \mu_{1} \alpha_{2}+4\alpha_{2}^{2}
\cos^{2}\xi + \mu_{1}^{2}\big/ 2)  \, ,
\end{split}   \label{t1:NR}  \\[6pt]
\begin{split}
0 =&~ 8 \sin\xi ~ \ddot{\theta}_{2}+ 16 \cos\xi ~\dot{\xi}\dot{\theta}_{2}
+\sin\xi \cos^{2}\xi \sin\theta_{2} \Big(16 \mu_{2} \alpha_{6} - 8 \mu_{2}
\alpha_{4} \cos\theta_{2}+ 8 \mu_{2} \alpha_{2} \cos\theta_{1} 
\\
&~ - \qty(\mu_{2} + 8 \alpha_{4}) \qty(2 \mu_{3} - \mu_{1} \cos\theta_{1}
+\mu_{2} \cos\theta_{2}) - 8 \qty(-2 \alpha_{4} \alpha_{6} - \alpha_{2}
\alpha_{4} \cos\theta_{1}) \Big) 
\\
&~ + \sin\xi \sin 2\theta_{2} \qty(8 \mu_{2} \alpha_{4} + 4 \alpha_{4}^{2}
\sin^{2}\xi + \mu_{2}^{2}\big/ 2)  \, ,
\end{split}   \label{t2:NR}  \\[6pt]
\begin{split}
0 =&~ \ddot{\xi}+\sin 2\xi \qty(\dot{\theta}_{1}^{2}-\dot{\theta}_{2}^{2})
+T_{\xi} \, ,
\end{split}  \label{xi:NR}
\end{alignat}
\end{subequations}
%%%%

where
%%%%
\begin{align}
\begin{split}
T_{\xi} =&~ 2 \qty(2 \mu_{3} - \mu_{1} \cos\theta_{1} + \mu_{2} \cos\theta_{2})
\Big(-1 - \frac{\alpha_{6}}{2} - \frac{\alpha_{2}}{2} \cos\theta_{1}
+\frac{\alpha_{4}}{2} \cos\theta_{2}
\\
&~ + \qty(2 \mu_{3} - \mu_{1} \cos\theta_{1}+ \mu_{2} \cos\theta_{2})\big/ 32
\Big) \sin 4\xi + 2 \sin 4\xi ~ \Big( \alpha_{6}^{2} - \alpha_{4} \alpha_{6}
\cos\theta_{2}
\\
&~ - \frac{1}{2} \cos\theta_{1} \qty(- 2\alpha_{2} \alpha_{6} +\alpha_{2}
\alpha_{4} \cos\theta_{2}) \Big) + \sin 2\xi ~ \Bigg\{ \frac{1}{2} \Big( \alpha_{2}^{2}
\cos^{2}\theta_{1} \cos^{2}\xi - \alpha_{4}^{2} \cos^{2}\theta_{2} \sin^{2}\xi
\Big) 
\\
&~ -\frac{1}{16} \qty(\mu_{1}^{2} \sin^{2}\theta_{1} - \mu_{2}^{2}
\sin^{2}\theta_{2}) - \qty(\mu_{1} \alpha_{2} \sin^{2}\theta_{1} - \mu_{2}
\alpha_{4} \sin^{2}\theta_{2})
\\
&~ - \frac{\alpha_{2}^{2}}{2} \Big( \sin^{2}\theta_{1}+\sin^{2}\xi \cos^{2}
\theta_{1} \Big) + \frac{\alpha_{4}^{2}}{2} \Big( \sin^{2}\theta_{2}+\cos^{2}
\xi \cos^{2}\theta_{2} \Big) \Bigg\}  \, .
\end{split}
\end{align}
%%%%

Using (\ref{act:NRa}) we next calculate the momenta conjugate to the isometry coordinates
as
%%%%
\begin{equation}\label{conmom:NR}
E \equiv \frac{\partial L_{P}}{\partial \dot{x}_{+}}
= -\frac{1}{4} \mathcal{M}\dot{x}_{+}  \, , 
\qquad P_{\Phi_{i}}\equiv\frac{\partial L_{P}}{\partial
\dot{\Phi_{i}}} =0 \, , \quad \qty(\Phi_{i} =\{\phi_{1}
\phi_{2},\psi \})  \, .
\end{equation}
%%%%

Using (\ref{chgJ:BD}) and (\ref{eom:xp}) it is easy to check that the corresponding charges ($J$)
are conserved:
%%%%
\begin{equation}\label{ConChg:NR}
\partial_{\tau}E=0 \,  \quad (\text{on-shell})\,, \qquad
\partial_{\tau}P_{\Phi_{i}}=0 \, .
\end{equation}
%%%%

We may now compute the energy-momentum tensors using the definition (\ref{EM:gen}). It is
easy to check that\footnote{Notice that, in order to avoid clutter in the resulting expressions
\textendash{} which are rather large \textendash{} from here on we set 
$\mu_{1}=\mu_{2}=\mu_{3}=\mu$.}
%%%%
\begin{align}
%\begin{split}
\partial_{\tau}T_{\tau\tau} &= \mathcal{R}   \, ,
\label{dervir1:NR} \\[6pt] 
\partial_{\tau}T_{\tau\sigma} &= 0  \, ,
\label{dervir2:NR}
%\end{split}
\end{align}
%%%%
where 
%%%%
\begin{align}\label{VirCon:NR}
\begin{split}
\mathcal{R} =&~ \frac{\mu}{64}\Bigg[ -\cos^{2}\xi \sin\theta_{1} ~ \dot{\theta}_{1}
\Big\{ 4\mu + 2\qty(\mu - 4) \cos\theta_{1} -4\mu \cos 2\xi 
\\
&~ -\qty(4-\mu) \Big( \cos\qty(\theta_{1}-2\xi) +\cos\qty(\theta_{1}+2\xi)
-\cos\qty(\theta_{2}-2\xi) - \cos\qty(\theta_{2}+2\xi) \Big)\Big\}
\\
&~ +\sin^{2}\xi ~ \dot{\theta}_{2} \Big\{ -4 \qty(-2\mu +\qty(\mu -4)\cos\theta_{1}
+4\cos\theta_{2}) \cos^{2}\xi \sin\theta_{2}
\\
&~ -2\sin 2\theta_{2} \qty(-4+\mu \sin^{2}\xi) \Big\} - \sin\xi \cos\xi ~ \dot{\xi}
\Big\{ 2\qty(\mu - 4)\qty(\cos 2\theta_{1} -\cos 2\theta_{2})
\\
&~ + 4\qty(-2+\cos\theta_{1}-\cos\theta_{2}) \Big( -2\qty(\mu+4)+\qty(\mu - 4)
\qty(\cos\theta_{1} -\cos\theta_{2}) \Big) \cos 2\xi \Big\} \Bigg]  \, .
\end{split}
\end{align}
%%%%

However, the Virasoro consistency conditions $\displaystyle \partial_{\tau}T_{ab}=0$ require
us to set $\mathcal{R}=0$ in (\ref{VirCon:NR}). Using this latter requirement and (\ref{tCon:NR})
we may now solve for $\dot{\xi}$ algebraically and substitute the resulting solution into the eoms
(\ref{t1:NR}) and (\ref{t2:NR}) corresponding to $\theta_{1}$ and $\theta_{2}$,
respectively. The resulting eoms are obvious and we avoid writing them here. Moreover, if we
choose the $\theta_{2}$ invariant plane in the phase space described as
%%%%
\begin{equation}
\label{IP1:NR}
\theta_{2} \sim 0 \, ,   \qquad \Pi_{\theta_{2}}:=\dot{\theta}_{2}\sim 0  \, ,
\end{equation}
%%%%
then we observe that the resulting $\theta_{2}$ eom (\ref{t2:NR}) (after $\dot{\xi}$ substitution)
is satisfied trivially. The other two $\dot{\xi}$ substituted eoms (\ref{t1:NR}) and (\ref{xi:NR}) then 
reduce to
%%%%
\begin{subequations}\label{eom:NR}
\begin{alignat}{2}
\begin{split}
0 =&~ 8 \cos\xi ~ \ddot{\theta}_{1} +8 \sin 2\xi \sin\theta_{1}~
\dot{\theta_{1}}^{2} ~\mathscr{K}_{\theta_{1}} + \cos\xi \sin 2\theta_{1}
\qty(8\mu +4 \cos^{2}\xi+ \frac{\mu^{2}}{2}) 
\\
&~ -\cos\xi \sin^{2}\xi \sin 2\theta_{1} \Big( 8\mu \qty(\cos\theta_{1}
+2) +8\qty(1-\mu)+\mu \qty(-3 + \cos\theta_{1})\qty(\mu +8) \Big)  \, ,
\end{split}   \label{t12:NR}  \\[6pt]
\begin{split}
0 =&~ 4\ddot{\xi}+\frac{1}{2}\sin 2\xi \, \dot{\theta}_{1}^{2} +\mathscr{T}_{\xi} \, ,
\end{split}  \label{xi2:NR}
\end{alignat}
\end{subequations}
%%%%
where
%%%%
\begin{align}
\begin{split}
\mathscr{K}_{\theta_{1}} &=~ - \frac{\qty(16 + 6 \mu)\qty(1-\cos 2\xi)
+2\qty(\mu +8)\cos\theta_{1}+2\qty(\mu +8)\cos\theta_{1}\cos 2\xi)}
{2\sin\xi \Big(\qty(\mu +8) \qty(\cos 2\theta_{1}-1)+2\cos 2\xi
\qty(-3 +\cos\theta_{1})\qty(8-3\mu+\qty(\mu+8)\cos 2\xi)\Big)} \\
&~ +\frac{\qty(3+\cos\theta_{1}\cot^{2}\xi)\sin\xi}{\sin^{2}\theta_{1}
\qty(3-\cos\theta_{1})^{2}\cos 2\xi}  \, ,
\end{split}
\end{align}
%%%%
%%%%
\begin{align}
\begin{split}
\mathscr{T}_{\xi} =&~ \frac{\mu}{2}\qty(-3+\cos\theta_{1}) \sin 4\xi
+ \cos^{2}\theta_{1} \sin\xi \cos^{3}\xi
\\
&~ + 2\mu \Big(2+ \qty(-3+\cos\theta_{1}) \Big)\cos\theta_{1} \sin\xi
\cos^{3}\xi 
\\
&~ + \qty(\frac{\mu^{2}}{8}\qty(-3+\cos\theta_{1})^{2}+1)\cos^{3}
\xi \sin\xi - \mu \sin^{2}\theta_{1} \sin 2\xi
\\
&~ -\frac{1}{8}\cos\xi \Big\{ \qty(8+\mu^{2}) \sin^{2}\theta_{1}
\sin\xi +\sin^{3}\xi \Big( 8+9\mu^{2} -2\qty(-8+24 \mu +3\mu^{2})
\cos\theta_{1}
\\
&~ +\qty(8+16\mu +\mu^{2})\cos^{2}\theta_{1}  \Big) \Big\}  \, .
\end{split}
\end{align}
%%%%

In the next step, we make the following choice of the $\theta_{1}$ invariant plane in
the phase space:
%%%%
\begin{equation}
\label{IP2:NR}
\theta_{1} \sim 0 \, ,   \qquad \Pi_{\theta_{1}}:=\dot{\theta}_{1}\sim 0  \, ,
\end{equation}
%%%%
which clearly satisfies (\ref{t12:NR}). Subsequently, the $\xi$ eom (\ref{xi2:NR}) can be
written in the form
%%%%
\begin{equation}\label{xi3:NR}
\ddot{\xi} + \mathcal{A}_{\text{NR}}\sin 4\xi = 0  \, ,
\end{equation}
%%%%
where
%%%%
\begin{align}\label{ANR:NR}
\begin{split}
\mathcal{A}_{\text{NR}} = \frac{8 - 16 \mu + \mu^{2}}{32}  \, .
\end{split}
\end{align}
%%%%

Now from (\ref{tCon:NR}) and (\ref{VirCon:NR}) we notice that, for the successive choices of
the invariant planes in the phase space, namely (\ref{IP1:NR}) and (\ref{IP2:NR}), $\dot{\xi}
=0$. Moreover, this solution must be consistent with (\ref{xi3:NR}). Since $\displaystyle
0\leq \xi <\pi$, the possible solutions of (\ref{xi3:NR}) can be expressed as
%%%%
\begin{equation}\label{sol:xiNR}
\bar{\xi} =~ \frac{n \pi}{4} \, , \quad 0\leq n < 4\, , n\in \mathbb{Z} \, .
\end{equation}
%%%%

We now consider infinitesimal fluctuations ($\delta\theta_{1}\sim\eta$) around the
$\theta_{1}$ invariant plane. The resulting NVE can then be written as
%%%%
\begin{align}\label{etaNVE:NR}
\begin{split}
\ddot{\eta} - \frac{1}{8} \Big( 2\sin^{2}\bar{\xi} \qty(4-\mu^{2})
-8 \cos^{2}\bar{\xi} -16 \mu - \mu^{2} \Big) \eta \approx ~0  \, ,
\end{split}
\end{align}
%%%%
where $\bar{\xi}$ is given by (\ref{sol:xiNR}). Also notice that, in writing the above NVE
(\ref{etaNVE:NR}), we have neglected the second term in the R.H.S of (\ref{t12:NR}) since
this term is $\order{\eta^{3}}$.

Next, with the given solutions (\ref{sol:xiNR}), (\ref{etaNVE:NR}) can easily be solved. The
solutions can formally be written as
%%%%
\begin{equation}\label{etaSol:NR}
 \eta(\tau) =~ \mathsf{C}_{1} \cos\qty(\sqrt{\mathsf{C}_{0}}~ \tau)
 + \mathsf{C}_{2} \sin\qty(\sqrt{\mathsf{C}_{0}}~ \tau)  \, ,
\end{equation} 
%%%% 
where 
%%%%
\begin{equation}
  \mathsf{C}_{0} =
    \begin{cases}
      \frac{1}{8}\qty(\mu^{2}+16\mu+8)\, , & \text{for $n=0$} \\[3pt]
      \frac{1}{4}\qty(\mu^{2}+8\mu)\, , & \text{for $n=1$} \\[3pt]
      \frac{1}{8}\qty(3\mu^{2}+16\mu -8)\, , & \text{for $n=2$} \\[3pt]
      \frac{1}{4}\qty(\mu^{2}+8\mu)\, , & \text{for $n=3$} \, ,
    \end{cases}       
\end{equation}
%%%%
and $\mathsf{C}_{1}$ and $\mathsf{C}_{2}$ are constants of integration.

Clearly, these solutions (\ref{etaSol:NR}) are Liouvillian which reflect the underlying non-chaotic
dynamics of the string.

\subsubsection{Numerical results}\label{num:NR}
We now check the integrability of the string configuration numerically using the methodology
discussed in Appendix \ref{app:num}.

Using the embedding (\ref{ansatz:NR}) together with $\alpha_{2}=\alpha_{4}=\alpha_{6}
=\eta_{1}=1$, the resulting Hamilton's equations can be computed as\footnote{In order to perform
the numerical analysis, we choose to work with the original coordinates and set $x_{0}=t(\tau)$ and 
$x_{2}=\eta_{1}\tau$, with $\eta_{1}=1$.}
%%%%
\begin{subequations}
\begin{alignat}{4}
\begin{split}
    \dot{\theta_1}=&~ 4p_{\theta_1}\sec^2\xi  \, ,
\end{split} \label{Hamth:NR}   \\[5pt]
\begin{split}
	\dot{\xi}=&~ p_\xi   \, ,
\end{split}   \label{Hamxi:NR}   \\[5pt]
\begin{split}
    \dot{p_{\theta_1}}=&~ \frac{1}
		{128}\bigg(-16\cos^2\xi\sin(2\theta_1)+2\mu_1\cos^2\xi\sin\theta_1
		\big(\mu_1\cos\theta_1\cos^2\xi+\big(\mu_2+2\mu_3\big)\sin^2\xi\big)
		\\
		&+4\sin(2\theta_1)\sin^2(2\xi)+\mathcal{N}_3\bigg)  \, ,
\end{split}  \label{Hampth:NR}   \\[5pt]    
\begin{split}
    \dot{p_\xi}=&~ \frac{1}{128}\bigg(-32\cos^3\xi\sin\xi-32\cos^2\theta_1\cos^3\xi\sin\xi	
+2(\mu_2-\mu_1\cos\theta_1)\big(\mu_2+4\mu_3-\mu_1\cos\theta_1\big)
\\
&\times\cos^3\xi\sin\xi+32\cos\xi\sin^3\xi+32\cos^2\theta_1\cos\xi\sin^3\xi-2\cos\xi
\sin\xi\big(\mu_1^2\sin^2\theta_1   
\\
&+\big(\mu_2-\mu_1\cos\theta_1\big)
\big(\mu_2+4\mu_3-\mu_1\cos\theta_1\big)\sin^2\xi\big)+\mathcal{N}_4
-512 p_{\theta_1}^2\sec^2\xi\tan\xi+2\mu_3^2\sin(4\xi)\bigg) \, ,
\end{split}   \label{Hampxi:NR} 
\end{alignat}
\end{subequations}
%%%%
where the detailed expressions for the functions $\mathcal{N}_3$ and $\mathcal{N}_4$ are
given in Appendix \ref{der:NR}.

\begin{figure}[h!]
\begin{multicols}{2}
    \includegraphics[width=\linewidth]{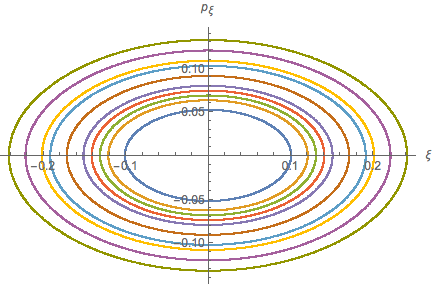}\par 
    \includegraphics[width=\linewidth]{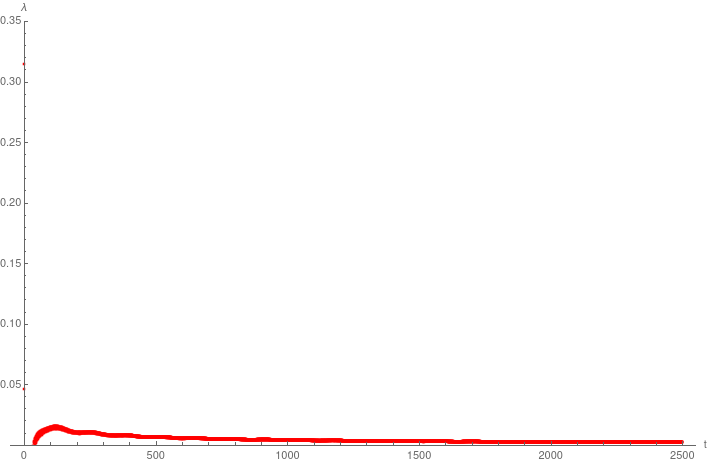}\par 
    \end{multicols}
\begin{multicols}{2}
    \includegraphics[width=\linewidth]{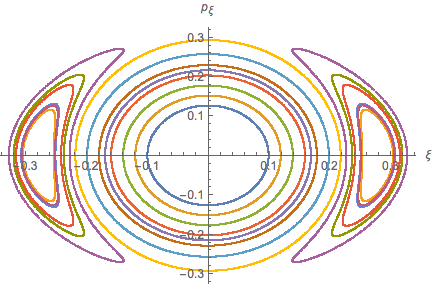}\par
    \includegraphics[width=\linewidth]{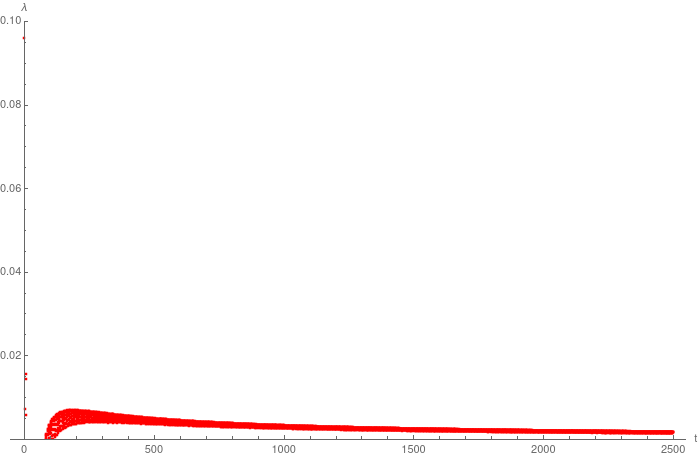}\par
\end{multicols}
\caption{Numerical plots of the Poincar\'{e} sections (\textbf{Left column}) and
Lyapunov exponents (\textbf{Right column}) for non-relativistic ABJM. Here
we set the energy of the string $E_{0}=0.3$. The top plots are for
$\mu_{1}=\mu_{2}=\mu_{3}=0.01$ while the bottom plots are for 
$\mu_{1}=\mu_{2}=\mu_{3}=0.8$. The Poincar\'{e} sections are undistorted
foliations of KAM tori in the phase space and for large time $t$ the Lyapunov
exponent decays to zero. These are indications of the non-chaotic dynamics of the
string configuration.}
\label{fig4:NR}
\end{figure}

\begin{figure}[h]
	\begin{multicols}{2}
		\includegraphics[width=\linewidth]{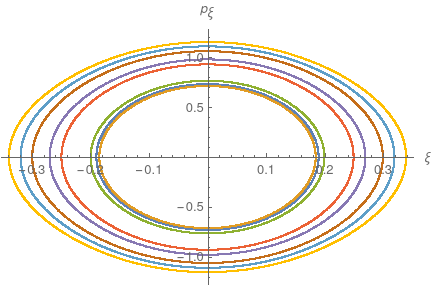}\par 
		\includegraphics[width=\linewidth]{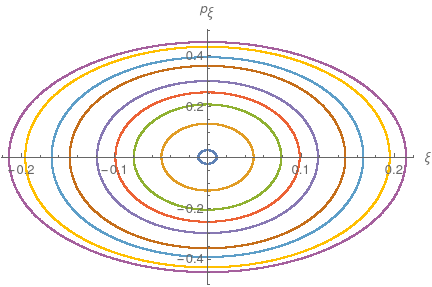}\par 
	\end{multicols}
	\caption{Additional plots of the Poincar\'{e} sections for non-relativistic ABJM. On the left plot we set $\mathbf{E_{0}=0.95}$ and $ \mu_{1}=\mu_{2}=\mu_{3}=~0.55 $, and on the right plot we set  $\mathbf{E_{0}=0.55}$, $ \mu_{1}=\mu_{2}=\mu_{3}=~0.1 $.}
	\label{Fig:New4:NR}
\end{figure}

As in our previous cases, we set $ \theta_2 =p_{\theta_2}=0 $ throughout the rest of the
analysis. The Poincar\'{e} sections, plotted in the left column of Fig.\ref{fig4:NR}, are obtained
by setting $ E=E_0=0.3 $ (The plots corresponding to other values of the energy $ E=0.95 $, $ E=0.55 $ are shown in Fig.\ref{Fig:New4:NR}). We as well set the following values of the YB deformation parameters:
$\mu_1=\mu_2=\mu_3=0.01$, $ 0.8 $. On top of that, the initial conditions are chosen as
$\theta_{1}(0)=0.1$ and $p_\xi(0)=0$. The random data set for $p_{\theta_{1}}(0)$ is then
generated by choosing $ \xi (0) \in [0, 1] $. The Poincar\'{e} section is obtained by collecting the
data set $ \lbrace \xi, p_\xi\rbrace $ every time the orbits pass through the $ \theta_1=0 $
hyper-plane.

In order to calculate the Lyapunov exponent ($\lambda$), plotted in the right column in
Fig.\ref{fig4:NR}), the corresponding initial conditions are chosen as $ \lbrace \theta_1(0)=0.1, 
\xi(0)=0.1, p_\xi(0)=0, p_{\theta_1}(0)=0.159 \rbrace $ such that the Hamiltonian constraints 
(\ref{Hamil}), (\ref{app:Vir}) are satisfied. The initial separation between the orbits as defined
in (\ref{app:Lyav}) is fixed at $ \Delta X_{0}=10^{-7} $ as before. This finally yield a zero
Lyapunov exponent at large time, like in the previous three examples. This shows a non-chaotic
motion for the dynamical phase space under consideration. For YB parameter value $ 0.8 $, the
initial conditions are set to be $ \lbrace \theta_1(0)=0.1, \xi(0)=0.256, p_\xi (0)=0,
p_{\theta_1}(0)=0.093 \rbrace $. Thus we find consistency between the analytical and the
numerical analyses and the string indeed undergoes non-chaotic dynamics.

%%%%
%%%%

%%%%
%%%%

%%%%%%%%%%%%%%%%%%%%%%%%%%%%%%%%%%%%%%%%%%%
\section{Final remarks and future directions}\label{Conclusion}
We confirm the non-chaotic dynamics for a class of Yang-Baxter (YB) deformed $ \text{AdS}_{4}
\times \text{CP}^{3} $ (super) string sigma models.
The deformed backgrounds that we considered in our analysis are in fact dual to various
deformations of the ABJM model \cite{Aharony:2008ug} at strong coupling
\cite{Negron:2018btz}-\cite{Rado:2021mji}, \cite{Imeroni:2008cr}. These backgrounds
are generated through the Yang-Baxter (YB) deformations: there exist classical
$r$-matrices that satisfy classical YB equation \cite{Klimcik:2002zj,Klimcik:2008eq}. %Interestingly, these deformed
%backgrounds can be generated by a TsT transformation \cite{Imeroni:2008cr} indicating that
%they are indeed solutions to the type IIA supergravity.

Interestingly, the YB deformed
backgrounds can be generated by a TsT transformation \cite{Imeroni:2008cr} on $AdS_4\times CP^3$ background with real deformation parameters. Our analysis reveals the absence of non-integrability for the given string embedding, which is consistent with the aforementioned fact as well as the analysis done by authors in \cite{Frolov:2005dj}  for the real $\beta$-deformation of $\mathcal{N}$=4 SYM. On the other hand, one loses integrability for the complex deformation parameter \cite{Giataganas:2013dha}.

\vspace{5pt}

The primary motivation for our study stemmed from the absence of any systematic analysis
of the integrable structures of these class of deformed backgrounds. This is in stark contrast
to the undeformed $\text{AdS}_{4}\times \text{CP}^{3}$ case where both analytical and
numerical confirmations of the integrability of string sigma models have been established
\cite{Arutyunov:2008if}-\cite{Sorokin:2010wn}.

\vspace{5pt}

In our investigation, we have used both analytical as well as numerical methods. For our analytical 
computations, we have used the famed Kovacic's algorithm
\cite{Kovacic:1986,Saunders:1981,Kovacic:2005,Nunez:2018qcj} which checks the Liouvillian
(non-)integrability of linear homogeneous second order ordinary differential equations of the form
(\ref{Kov:Dif}) via a set of \textit{necessary but not sufficient} criteria. In our analysis, we have been able
to recast the dynamical equations of motion of the propagating string in the form of (\ref{Kov:Dif})
and checked the fulfilment of the criteria put forward by the algorithm. This established the
non-chaotic dynamics of the string in the corresponding deformed backgrounds. 

\vspace{5pt}

Our analytical results have been substantiated by numerical analysis where we estimated various
chaos indicators of the theory, namely, the Poincar\'{e} section and the Lyapunov exponent. In
our computations, using the standard Hamiltonian formulation 
\cite{Zayas:2010}-\cite{Banerjee:2018ifm}, we explicitly checked that the shapes of the KAM
tori never get distorted as we increase the YB deformation parameters in all the four cases. At this point, we must mention that the nice foliations of the Poincar\'{e} sections that we observed in Figs.\ref{fig1:BD},\ref{fig2:NC},\ref{fig3:DD},\ref{fig4:NR} do not necessarily guarantee that the system is non-chaotic for the entire range of values of the parameters in the theories \textemdash{} string energy ($ E $) and various Yang-Baxter deformation parameters \textemdash{} as was observed earlier, e.g., in \cite{Basu:2012,Giataganas:2017guj}. In order to establish our claim, we take additional set of values of the above mentioned parameters and find that our reults are indeed consistent. The corresponding plots are given in Figs.\ref{Fig:New1:BD},\ref{Fig:New2:NC},\ref{Fig:New3:DD},\ref{Fig:New4:NR}. Also,
the Lyapunov exponent decays to zero with time. These two results allow us to conclude that
the phase space of the propagating string does not show any chaotic behaviour, thereby
establishing consistency with our analytical results. \textit{Nevertheless, our analyses do not prove
integrability following the traditional Lax pair formulation; rather, it disproves non-integrable
structure for certain physical stringy configurations.}

\vspace{5pt}

The (semi)classical strings, those probe these YB deformed backgrounds, are dual to a class of
single trace operators in some sub-sector(s) of these deformed ABJM models. Our analysis,
therefore points towards an underlying integrable structure associated with these deformed ABJM
models. A systematic analysis of the Lax pairs would further strengthen this claim.

\vspace{5pt}

From the perspective of the deformed ABJMs, a similar investigation on the dilatation operators
should shed further light on an integrable structure associated with the dual quantum field theory.
This would be an interesting future direction to look for which would eventually take us into a new
class of Gauge/String dualities those are associated with an underlying integrable structure.
%%%%%%%%%%%%%%%%%%%%%%%%%%%%%%%%%%%%%%%%%
\section*{Acknowledgments}
J.P., H.R. and D.R. are indebted to the authorities of IIT Roorkee for their unconditional support
towards researches in basic sciences. D.R. would also like to acknowledge The Royal Society, UK
for financial assistance, and acknowledges the Grant (No. SRG/2020/000088), and Mathematical Research Impact Centric Support (MATRICS) grant (MTR/2023/000005) received from the
Science and Engineering Research Board (SERB), Govt. of India. AL would like to thank the authorities of IIT
Madras and IOP Bhubaneswar for supporting research in fundamental physics. He also acknowledges the financial
support from the project ``Quantum Information Theory'' (No. SB20210807PHMHRD008128).
%%%%%%%%%%%%%%%%%%%%%%%%%%%%%%%%%%%
\appendix
\numberwithin{equation}{section}
\renewcommand{\theequation}{\thesection\arabic{equation}}
%%%%%%%%%%%%%%%%%%%%%%%%%%%%%%%%%%%%%%%%%
\section{The Kovacic's algorithm}\label{Kova}

The Kovacic's algorithm is a systematic method to determine whether a second-order linear
homogeneous differential equation of the form
%%%%
\begin{equation}
\label{Kov:Dif}
\eta''(z)+M(z)\eta'(z)+N(z)\eta(z)=0 \, ,
\end{equation}
%%%%
where $M(z)$, $N(z)$ are polynomial coefficients, are integrable in the Liouvillian sense. This
implies the existence of the solutions of (\ref{Kov:Dif}) in the form of algebraic functions,
trigonometric functions and exponentials. 

We here discuss only the necessary details regarding the formalism as the detailed mathematical
analysis is rather involved. One wishes to find the relation among $M(z)$, $M'(z)$ and $N(z)$
that makes the DE (\ref{Kov:Dif}) integrable. In order to achieve this, we start from the change
in variable of the form
%%%%
\begin{equation}\label{Kov:Var}
\eta(z) = \exp \qty[\int \dd z \qty(w(z)-\frac{M(z)}{2})] \, .
\end{equation}
%%%%

Eq.(\ref{Kov:Var}) permits us to express (\ref{Kov:Dif}) in the following form:
%%%%
\begin{equation}\label{Kov:Fnl}
w'(z) +w^{2}(z) =\mathcal{V}(z)=\frac{2M'(z)+M^{2}(z)-4N(z)}{4} \, .
\end{equation}
%%%%

Now the group of symmetry transformations, $\mathcal{G}$, of the solutions of the DE
(\ref{Kov:Dif}) is a subgroup of $SL\qty(2,\mathbb{C})$: $\mathcal{G}\subset
SL\qty(2,\mathbb{C})$. The following four cases are of interest
\cite{Kovacic:2005,Nunez:2018qcj}:

%%%%
\begin{enumerate}[(i)]

\item The subgroup is generated by 
		%%%%
		\begin{equation*}
			\mathcal{G} = \mqty*(a & 0 \\ b & 1/a) \, ,
			\quad a, b \in \mathbb{C} \, .
		\end{equation*}
		%%%%
		In this case $w(z)$ is a rational function of degree $1$.
		
\item The subgroup is generated by 
		%%%%
		\begin{equation*}
			\mathcal{G} = \mqty*(c & 0 \\ 0 & 1/c) \, ,
			\quad
			\mathcal{G} = \mqty*(0 & c \\ -1/c & 0) \, ,
			\quad
			c \in \mathbb{C} \, .
		\end{equation*}
		%%%%
		In this case $w(z)$ is a rational function of degree $2$.
		
\item $\mathcal{G}$ is a finite group, excluding the above two possibilities. In this case $w(z)$
is a rational function of degree either $4$, $6$, or $12$.

\item The group $\mathcal{G}$ is $SL(2,\mathbb{C})$. If the solution $w(z)$ at all exists, they
non-Liouvillian.		

\end{enumerate}
%%%%

There exists a set of three necessary but not sufficient conditions for the rational polynomial
function $\mathcal{V}(z)$ which are compatible with the above group theoretic analysis. These
can be enumerated as follows \cite{Kovacic:1986,Saunders:1981}:

%%%%
\begin{enumerate}[\textbf{Cd.}(i)]

\item $\mathcal{V}(z)$ has pole of order $1$, or $2n$ $(n \in \mathbb{Z}^{+})$. Also, the
order of $\mathcal{V}(z)$ at infinity\footnote{Here we define the order at infinity of a polynomial
as the \emph{difference} between the highest power of its argument in the denominator
and that in the numerator. This convention is different from that used in \cite{Nunez:2018qcj}
where the \emph{difference} is replaced by \emph{subtraction}.} is either $2n$ or greater than
$2$.

\item $\mathcal{V}(z)$ either has pole of order $2$, or poles of order $2n + 1$ greater than
$2$.

\item $\mathcal{V}(z)$ has poles not greater than $2$ and the order of $\mathcal{V}(z)$ at
infinity is at least $2$.

\end{enumerate}
%%%%
If any one of these criteria is satisfied, we are eligible to apply the Kovacic's algorithm to the DE
(\ref{Kov:Dif}). We then need to determine whether $w(z)$ is a polynomial function of degree
$1$, $2$, $4$, $6$, or $12$ in which case (\ref{Kov:Dif}) turns out to be integrable. On the
contrary, if none of the above criteria is satisfied, the solution to (\ref{Kov:Dif}) is
non-Liouvillian and ensures the non-integrability of the DE (\ref{Kov:Dif}).

%%%%%%%%%%%%%%%%%%%%%%%%%%%%%%%%%%%%%%%%%
\section{Numerical Methodology}\label{app:num}

In the present work, we focus on two chaos indicators namely, the Poincar\'{e} section and
the Lyapunov exponent \cite{Zayas:2010}-\cite{Basu:2011b}. For the familiarity of the
reader, below we briefly elaborate on them and outline basic steps to calculate these entities
in a holographic set up.

The signatures of integrability or non-integrability can be differentiated by looking into the
phase space dynamics of the system. Integrable systems do not exhibit chaos and the
trajectories are (quasi)periodic at equilibrium points. Non-integrable systems, on the other
hand, are associated with the phase space that could be mixed showing (quasi)periodic
orbits for some initial conditions and chaotic for others. 

\iffalse
\begin{figure}
    \centering
    \includegraphics[height=6.5in,width=6.5in]{fig1.pdf}
    \caption{ A. Poincaré sections for (1(a), 2(a)) $\beta$ deformed ABJM with $E=0.01$, (1(b), 2(b)) Noncommutative ABJM with $E=0.4$, (1(c), 2(c)) Dipole deformed ABJM with $E=0.35$ and (1(d), 2(d)) Non-Relativistic ABJM with $E=0.3$. Each of these Poincare sections exhibit a nice foliation for two different choices of the YB parameter those are ($ 0.01 $ and $ 0.8 $) depicted in the figures showing an integrable structure for the associated phase space dynamics. B. Lyapunov exponents for (3(a), 4(a)) $\beta$ deformed ABJM with $E=0.01$, (3(b), 4(b)) Noncommutative ABJM with $E=0.4$, (3(c), 4(c)) Dipole deformed ABJM with $E=0.35$ and (3(d), 4(d)) Non-Relativistic ABJM with $E=0.3$. In each of these examples, the Lyapunov approaches zero at late times which reveals an integrable structure for the phase space dynamics. As our analysis reveals, the qualitative nature of these plots remains the same for larger deformations.}
    \label{fig1}
\end{figure}
\fi

For a $ 2N $ dimensional integrable phase space, there are $ N $ conserved charges $ Q_i $,
those define an $ N $ dimensional hypersurface in the phase space known as the KAM tori.
For such systems, the phase space trajectory flows are complete and they appear with a
nicely foliated picture of the phase space. Different initial conditions give rise to different sets
of trajectories in the phase space those are in the form of the tori. In numerical investigations,
Poincar\'{e} sections\footnote{It is a lower dimensional slicing hypersurface of an $ N $
dimensional foliated KAM tori.} (see left panels of Figs. \ref{fig1:BD},\ref{fig2:NC},
\ref{fig3:DD},\ref{fig4:NR}) are essentially the footprints of such foliations in the phase space
\cite{Basu:2011a}. As the strength of the non-integrable deformation increases, most of these
tori get destroyed and one essentially runs away from the foliation picture. This results into a
chaotic motion and Poincar\'{e} sections loose its structure, eventually becoming like a random
distribution of points in the phase space.

%\begin{figure}
    %\centering
    %\includegraphics[height=4in,width=4in]{fig2.pdf}
    %\caption{(a) Lyapunov exponents for (a) $\beta$ deformed ABJM with $E=0.01$, (b) Noncommutative ABJM with $E=0.4$, (c) Dipole deformed ABJM with $E=0.4$ and (d) Non-Relativistic ABJM with $E=0.6$. In each of these examples, the Lyapunov approaches zero at late times which reveals an integrable structure for the phase space dynamics.}
   % \label{fig2}
%\end{figure}

Lyapunov exponents (see right panels of Figs. \ref{fig1:BD},\ref{fig2:NC},\ref{fig3:DD},
\ref{fig4:NR}), on the other hand, are the signature trademarks of a chaotic motion. They
encode the sensitivity of the phase space trajectories on the initial conditions and are defined as\footnote{For a $ 2N $ dimensional phase space, there are in principle $ 2N $ Lyapunov
exponents satisfying the constraint, $\sum_{i=1}^{2N}\lambda_i=0 $. In this paper,
however, we compute the largest positive Lyapunov among all these possible ones.}\cite{Zayas:2010}-\cite{Basu:2011b}
%%%%
\begin{equation}\label{app:Lyav}
\lambda = \lim_{t \rightarrow \infty}\lim_{\Delta X_0 \rightarrow 0}
\frac{1}{t}\log \frac{\Delta X (X_0 , t)}{\Delta X (X_0 , 0)},
\end{equation}
%%%%
where, $ \Delta X $ is the infinitesimal separation between two trajectories in the phase space. For integrable trajectories, those pertaining to a particular KAM tori, the corresponding $ \lambda $ approaches zero at late times. On the other hand, it exhibits a nonzero value for chaotic orbits.

To calculate the above entities in a string theory set up, one has to start with the 2D string sigma
model description in (\ref{POL}). Given the conjugate momenta (\ref{con:mom}) and the
Hamiltonian (\ref{Hamil}), we study the corresponding Hamilton's equations of motion (for a given
string embedding) those are subjected to the Virasoro (or the Hamiltonian) constraints of the form 
\cite{Zayas:2010}
%%%%
\begin{align}
\label{app:Vir}
\begin{split}
\mathcal{H} &= T_{\tau \tau} \approx 0 \, ,  \\
T_{\tau \sigma} &= T_{\sigma \tau} \approx 0 \, .
\end{split}
\end{align}
%%%%

The above constraints (\ref{app:Vir}) are always satisfied during the time evolution of the
system. The initial data that satisfy (\ref{app:Vir}) are used to find solutions to the Hamilton's
equations of motion corresponding to different backgrounds those are listed above. These
solutions are what we call the \emph{phase space data} those are finally used to explore the
chaos indicators mentioned above.

%%%%%%%%%%
\section{Expressions for the coefficients in (\ref{eom:BD})}\label{cofT:BD}

\begin{align}
\begin{split}
T^{(1)}_{\theta_{1}} &= -4\alpha_{2}^{2}\qty(\alpha_{2}^{2}\sin^{2}\theta_{1}
\cos^{2}\xi + \cos^{2}\theta_{1}\sin^{2}2\xi \big/ 4) -4\sin^{2}\xi \cos^{2}\xi
\Big(\alpha_{2}^{2}\cos^{2}\theta_{1}+\alpha_{4}^{2}\cos^{2}\theta_{2}  
\\
&\quad +4\alpha_{2}\alpha_{6}\cos\theta_{1}-2\alpha_{2}\alpha_{4}\cos\theta_{1}
\cos\theta_{2}-4\alpha_{4}\alpha_{6}\cos\theta_{2} \Big) -\sin^{2}\theta_{1}
\sin^{2}\theta_{2}\sin^{4}2\xi \, \times    \\
&\quad \Big(\hat{\gamma}_{1}^{2}\alpha_{2}^{2}+2\alpha_{2}\alpha_{4}
\hat{\gamma}_{1}\hat{\gamma}_{2}+\hat{\gamma}_{2}^{2}\alpha_{4}^{2}
+2\alpha_{2}\alpha_{6}\hat{\gamma}_{1}\hat{\gamma}_{3}+2\alpha_{4}
\alpha_{6}\hat{\gamma}_{2}\hat{\gamma}_{3}+\hat{\gamma}_{3}^{2}
\alpha_{6}^{2} \Big) -4\alpha_{6}^{2}\sin^{2}2\xi \,.
\end{split}
\end{align}
%%%%

%%%%
\begin{align}
\begin{split}
T^{(2)}_{\theta_{1}} &= \alpha_{2}^{2}\qty(4\cos^{2}\xi -\sin^{2}2\xi
+\hat{\gamma}_{1}^{2}\sin^{2}\theta_{2}\sin^{4}2\xi)\sin 2\theta_{1}
+\qty(\alpha_{4}\hat{\gamma}_{2}+\alpha_{6}\hat{\gamma}_{3})^{2}
\sin 2\theta_{1} \sin^{2}\theta_{2} \times  \\
&\quad \sin^{4}2\xi +2\alpha_{2} \Big[ \alpha_{4}\qty(\sin\theta_{1}
\cos\theta_{2}\sin^{2}2\xi+\hat{\gamma}_{1}\hat{\gamma}_{2}\sin 2\theta_{1}
\sin^{2}\theta_{2}\sin^{4}2\xi )+\alpha_{6}\sin^{2}2\xi \, \times  \\
&\quad \qty(-2\sin\theta_{1}+\hat{\gamma}_{1}\hat{\gamma}_{3}
\sin 2\theta_{1}\sin^{2}\theta_{2}\sin^{2}2\xi) \Big] \, .
\end{split}
\end{align}
%%%%

%%%%
\begin{align}
\begin{split}
T^{(1)}_{\theta_{2}} &= -4\alpha_{2}^{2} \cos^{2}\xi\Big(\cos^{2}\theta_{1}
\sin^{2}\xi +\sin^{2}\theta_{1}\qty(1+4\hat{\gamma}_{1}^{2}\cos^{2}\xi
\sin^{4}\xi \sin^{2}\theta_{2})\Big) +8\alpha_{2}\alpha_{4}   \\
&\quad \cos^{2}\xi \sin^{2}\xi \qty(\cos\theta_{1}\cos\theta_{2}-\hat{\gamma}_{1}
\hat{\gamma}_{2}\sin^{2}\theta_{1}\sin^{2}\theta_{2}\sin^{2}2\xi)
-4\alpha_{4}^{2}\sin^{2}\xi \Big( \cos^{2}\theta_{2}\cos^{2}\xi   \\
&\quad +\sin^{2}\theta_{2}\qty(1+4\hat{\gamma}_{2}\sin^{2}\theta_{1}\sin^{2}\xi 
\cos^{4}\xi)\Big) -8\alpha_{2}\alpha_{6}\sin^{2}\xi \cos^{2}\xi \Big(2\cos\theta_{1}
+\hat{\gamma}_{1}\hat{\gamma}_{3}\sin^{2}\theta_{1}  \\
&\quad \sin^{2}\theta_{2}\sin^{2}2\xi \Big) +8\alpha_{4}\alpha_{6}\sin^{2}\xi
\cos^{2}\xi \qty(2\cos\theta_{2} -\hat{\gamma}_{2}\hat{\gamma}_{3}\sin^{2}
\theta_{1}\sin^{2}\theta_{2}\sin^{2}2\xi)    \\
&\quad -\alpha_{6}^{2} \sin^{2}2\xi\qty(4+\hat{\gamma}_{3}^{2}\sin^{2}\theta_{1}
\sin^{2}\theta_{2}\sin^{2}2\xi)  \, .
\end{split}
\end{align}
%%%%

%%%%
\begin{align}
\begin{split}
T^{(2)}_{\theta_{2}} &= -\qty(\hat{\gamma}_{1}^{2}\alpha_{2}^{2}
+2\hat{\gamma}_{1}\hat{\gamma}_{3}\alpha_{2}\alpha_{6}
+\hat{\gamma}_{3}^{2}\alpha_{6}^{2})\, \sin^{2}\theta_{1}
\sin2\theta_{2}\sin^{4}2\xi -2\alpha_{2}\alpha_{4}\sin^{2}2\xi \sin\theta_{2}
\\
&\quad \times \qty(\cos\theta_{1}+2\hat{\gamma}_{1}\hat{\gamma}_{2}
\sin^{2}2\xi\sin^{2}\theta_{1}\cos\theta_{2})-4\alpha_{4}^{2}\sin 2\theta_{2}
\sin^{4}\xi \qty(1 +4\hat{\gamma}_{2}^{2}\sin^{2}\theta_{1}\cos^{4}\xi)
\\
&\quad -2\alpha_{4}\alpha_{6}\sin^{2}2\xi \qty(2\sin\theta_{2}+\hat{\gamma}_{2}
\hat{\gamma}_{3}\sin^{2}\theta_{1}\sin 2\theta_{2}\sin^{2} 2\xi) \, .
\end{split}
\end{align}
%%%%

%%%%
\begin{align}
\begin{split}
T^{(1)}_{\xi} &=-\qty(\hat{\gamma}_{1}^{2}\alpha_{2}^{2}+2\hat{\gamma}_{1}
\hat{\gamma}_{2}\alpha_{2}\alpha_{4}+\hat{\gamma}_{2}^{2}\alpha_{4}^{2}
+\hat{\gamma}_{3}^{2}\alpha_{6}^{2}+2\hat{\gamma}_{1}\hat{\gamma}_{3}
\alpha_{2}\alpha_{6}+2\hat{\gamma}_{2}\hat{\gamma}_{3}\alpha_{4}\alpha_{6})
 \sin^{2}\theta_{1}\sin^{2}\theta_{2}\sin^{4}2\xi   \\
&\quad -4\alpha_{2}^{2}\sin^{2}\theta_{1}\cos^{2}\xi -4\alpha_{4}^{2}\sin^{2}
\theta_{2}\sin^{2}\xi +\sin^{2}2\xi \Big(-4 \alpha_{6}^{2}-\alpha_{2}^{2}\cos^{2} 
\theta_{1}-\alpha_{4}^{2}\cos^{2}\theta_{2}    \\
&\quad +2\alpha_{2}\alpha_{4}\cos\theta_{1} \cos\theta_{2}-4\alpha_{2}\alpha_{6}
\cos\theta_{1} +4\alpha_{4}\alpha_{6}\cos\theta_{2}\Big) \, .
\end{split}
\end{align}
%%%%

%%%%
\begin{align}
\begin{split}
T^{(2)}_{\xi} &= -2\Big(\alpha_{2}^{2} \qty(32\hat{\gamma}_{1}^{2}\sin^{3}\xi
\cos^{3}\xi \cos 2\xi \sin^{2}\theta_{1}\sin^{2}\theta_{2}-2\sin^{2}\theta_{1}
\sin 2\xi +\cos^{2}\theta_{1}\sin 4\xi)   \\
&\quad +2\alpha_{2} \sin 4\xi \Big[ \alpha_{4} \qty(-\cos\theta_{1}\cos\theta_{2}
+2\hat{\gamma}_{1}\hat{\gamma}_{2}\sin^{2}\theta_{1}\sin^{2}\theta_{2}
\sin^{2}2\xi)   \\
&\quad +2\alpha_{6} \qty(\cos\theta_{1}+\hat{\gamma}_{1}\hat{\gamma}_{3}
\sin^{2}\theta_{1}\sin^{2}\theta_{2}\sin^{2}2\xi) \Big] +2\sin 2\xi \Big[
\alpha_{4}^{2}\Big(\cos^{2}\theta_{2}\cos 2\xi +\sin^{2}\theta_{2}    \\
&\quad \times \qty(1+2\hat{\gamma}_{2}^{2}\sin^{2}2\xi \cos 2\xi
\sin^{2}\theta_{1})\Big) -4\alpha_{4}\alpha_{6}\cos2\xi \qty(\cos\theta_{2}
-\hat{\gamma}_{2}\hat{\gamma}_{3}\sin^{2}\theta_{1}\sin^{2}\theta_{2}
\sin^{2}2\xi)  \\
&\quad +2\alpha_{6}^{2}\cos2\xi \qty(2+\hat{\gamma}_{3}^{2}\sin^{2}
\theta_{1}\sin^{2}\theta_{2}\sin^{2}2\xi) \Big]\Big) \, .
\end{split}
\end{align}
%%%%

%%%%

%%%%%%%%%%
\section{Detailed expressions of $\mathcal{N}_{3}$ and $\mathcal{N}_{4}$ 
in (\ref{Hampth:NR}), (\ref{Hampxi:NR}) }\label{der:NR}
The expression for $\mathcal{N}_3$ in (\ref{Hampth:NR}) is given by 
	%%%%
	\begin{align}
	\mathcal{N}_3=-\frac{\mathcal{M}_1}{\mathcal{D}_1}-\frac{\mathcal{M}_2}{64\mathcal{D}_2^2}+\frac{\mathcal{M}_3}{2\mathcal{D}_1}-\frac{\mathcal{M}_4}{\mathcal{D}_1}+\frac{\mathcal{M}_5}{\mathcal{D}_3^2} \, ,
\end{align}
%%%%
where
%%%%
\begin{align*}
	\mathcal{M}_1&=\mu_1\cos^4\xi\sin\theta_1\bigg(-8\sqrt{2}+2\mu_2+4\mu_3
	+2\mu_1\cos\theta_1+\mu_1\cos(\theta_1-2\xi)\nonumber\\&+8\sqrt{2}\cos(2\xi)
	-2\mu_2\cos(2\xi)-4\mu_3\cos(2\xi)+\mu_1\cos(\theta_1+2\xi)\bigg) 
	\nonumber\\&\times\bigg(\mu_1^2\sin^2\theta_1+(\mu_2+2\mu_3-\mu_1
	\cos\theta_1)^2\sin^2\xi\bigg)  \, ,
\end{align*}
%%%%

%%%%
\begin{align*}
	\mathcal{D}_1&=	 	16+\cos^2\xi  \bigg[ 
	\mu_1^2\sin^2\theta_1+\bigg(\mu_2^2-2\mu_1(\mu_2+2\mu_3)\cos\theta_1
	+\mu^2_1\cos^2\theta_1\bigg)\sin^2\xi\bigg]+\mu_3(\mu_2
	\nonumber\\&+\mu_3)\sin^2(2\xi)  \, ,
\end{align*}
%%%%

%%%%
\begin{align*}
\mathcal{M}_2&=\mu_1\csc^4\xi\bigg(\big(2\big(\mu_2+2\mu_3\big)\sin\theta_1
+\mu_1\cot^2\xi\sin(2\theta_1)\big)\big(256E+4\mu_1^2\cos^2\xi\sin^2\theta_1
\nonumber\\
&+\big(\mu_2+2\mu_3-\mu_1\cos\theta_1\big)\big(-8\sqrt{2}+\mu_2+2\mu_3-
\mu_1\cos\theta_1\big)\sin^2(2\xi)\big)^2\bigg)  \, ,
\end{align*}
%%%%

%%%%
\begin{align*}
	\mathcal{D}_2=	\big(\mu_2+2\mu_3-\mu_1\cos\theta_1\big)^2\cos^2\xi
	+16\csc^2\xi	+\mu_1^2\cot^2\xi\sin^2\theta_1   \, ,
\end{align*}
%%%%

%%%%
\begin{align*}
	\mathcal{M}_3&=\mu_1\cos^2\xi\bigg(-8\sqrt{2}+2\mu_2+4\mu_3+2\mu_1
	\cos\theta_1+\mu_1\cos(\theta_1-2\xi)+8\sqrt{2}\cos(2\xi)
	\nonumber\\
	&-2\mu_2\cos(2\xi)-4\mu_3\cos(2\xi)+\mu_1\cos(\theta_1+2\xi)\bigg)\sin\theta_1
	\bigg(128 E
	\nonumber\\
	&+2\mu_1^2\cos^2\xi\big(\sin^2\theta_1+\cos^2\theta_1
	\sin^2\xi \big)-\frac{1}{2}\big(\big(8\sqrt{2}-\mu_2-2\mu_3\big)(\mu_2+2\mu_3)
	\nonumber\\
	&+2\mu_1\big(-4\sqrt{2}+\mu_2+2\mu_3\big)
	\cos\theta_1\big)\sin^2(2\xi)\bigg)   \, ,
\end{align*}
%%%%

%%%%
\begin{align*}
	\mathcal{M}_4=&	\mu_1\cos^2\xi\bigg(\mu_1\cos^2\xi\sin(2\theta_1)
	+2(\mu_2+2\mu_3)	\sin\theta_1\sin^2\xi\bigg)\times\bigg(128E
	\nonumber\\
	&+2\mu_1^2\cos^2\xi\big(\sin^2\theta_1+\cos^2\theta_1\sin^2\xi \big)
	-\frac{1}{2}\big(\big(8\sqrt{2}-\mu_2-2\mu_3\big)
	(\mu_2+2\mu_3)
	\nonumber\\
	&+2\mu_1\big(-4\sqrt{2}+\mu_2+2\mu_3\big)\cos\theta_1\big)\sin^2(2\xi)
	\bigg)    \,  ,
\end{align*}
%%%%

%%%%
\begin{align*}
\mathcal{M}_5&=	
\mu_1\cos^4\xi\big(\mu_2+2\mu_3+\mu_1\cos\theta_1\cot^2\xi\big)
\bigg(\mu_1^2\cot^2\xi-2\mu_1(\mu_2+2\mu_3)\cot\theta_1\csc\theta_1
\nonumber\\
&+\big(\mu_2+2\mu_3\big)^2\csc^2\theta_1+\mu_1^2\csc^2\xi\bigg)
\sin\theta_1\bigg(4\mu_1^2\cos^2\xi+\mu_1^2\cot^2\theta_1\sin^2(2\xi)
\nonumber\\
&+2\mu_1\big(4\sqrt{2}-\mu_2-2\mu_3\big)\cot\theta_1\csc\theta_1\sin^2(2\xi)
+\csc^2\theta_1\big(256 E
\nonumber\\
&-\big(8\sqrt{2}-\mu_2-2\mu_3\big)(\mu_2+2\mu_3)\sin^2(2\xi)\big)\bigg)  \, ,
\end{align*}
%%%%

%%%%
\begin{align*}
	\mathcal{D}_3=\mu_1^2\cot^2\xi+\big(\mu_2+2\mu_3-\mu_1\cos\theta_1\big)^2
	\cos^2\xi\csc^2\theta_1+16\csc^2\theta_1\csc^2\xi  \, .
\end{align*}
%%%%

The expression for $\mathcal{N}_4$ in (\ref{Hampxi:NR}) is given by 
%%%%
\begin{align}
	\mathcal{N}_4=\frac{\mathcal{M}_6}{2\mathcal{D}_2}+\frac{\mathcal{M}_7}
	{\mathcal{D}_1}+\frac{\mathcal{M}_8}{\mathcal{D}_1^2}-\frac{\mathcal{M}_9}
	{\mathcal{D}_1^2}-\frac{\mathcal{M}_{10}}{\mathcal{D}_1}+\frac{2\mathcal{M}_{11}}
	{\mathcal{D}_1}  \, ,
\end{align}
%%%%
where
%%%%
\begin{align*}
	\mathcal{M}_6&=16\sin^2\theta_1\sin(2\xi)\big(\mu_2+2\mu_3-\mu_1\cos\theta_1
	\big)^2\csc^2\xi\bigg(\big(\mu_2+2\mu_3-\mu_1\cos\theta_1\big)\big(-8\sqrt{2}+\mu_2
	\nonumber\\
	&+2\mu_3-\mu_1\cos\theta_1\big)\cos^2\xi+64E\csc^2\xi+\mu_1^2\cot^2\xi\sin^2\theta_1
	\bigg)\sin^3(2\xi)   \, ,
\end{align*}
%%%%

%%%%
\begin{align*}
	\mathcal{M}_7&=\bigg(\mu_1^2\sin^2\theta_1+\big(\mu_2+2\mu_3-\mu_1\cos\theta_1
	\big)^2\sin^2\xi\bigg)\sin(2\xi)\bigg(128E
	\nonumber\\
	&+2\mu_1^2\cos^2\xi\big(\sin^2\theta_1+\cos^2\theta_1\sin^2\xi \big)-\frac{1}
	{2}\big(\big(8\sqrt{2}-\mu_2-2\mu_3\big)(\mu_2+2\mu_3)
	\nonumber\\
	&+2\mu_1\big(-4\sqrt{2}+\mu_2+2\mu_3\big)\cos\theta_1\big)\sin^2(2\xi)\bigg)  \, ,
\end{align*}
%%%%

%%%%
\begin{align*}
\mathcal{M}_8&=	
2\mu_3^2\sin(4\xi)+\bigg(2\cos^2\xi\big(\mu_1^2\sin^2\theta_1+\big(\mu_2+2\mu_3
-\mu_1\cos\theta_1\big)^2\sin^2(\xi)\big)\bigg)
\nonumber\\
&\times\bigg(64E+\cos^2\xi\big(\mu_1^2\sin^2\theta_1+\big(\mu_2-\mu_1\cos\theta_1
\big)\big(\mu_2+4\mu_3-\mu_1\cos\theta_1\big)\sin^2\xi\big)+\mu_3^2\sin^2(2\xi)
\nonumber\\
&-2\sqrt{2}\big(\mu_2+2\mu_3-\mu_1\cos\theta_1\big)\sin^2(2\xi)\bigg)\bigg(2\big(\mu_2
-\mu_1\cos\theta_1\big)\big(\mu_2+4\mu_3-\mu_1\cos\theta_1\big)\cos^3\xi\sin\xi
\nonumber\\
&-2\cos\xi\sin\xi  
\big(\mu_1^2\sin^2\theta_1+\big(\mu_2-\mu_1\cos\theta_1\big)\times\big(\mu_2+4\mu_3
-\mu_1\cos\theta_1\big)\sin^2\xi\big)  + 2\mu_3^2\sin(4\xi) \bigg)  \, ,
\end{align*}
%%%%

%%%%
\begin{align*}
	\mathcal{M}_9&=\bigg(64E+\cos^2\xi\big(\mu_1^2\sin^2\theta_1+\big(\mu_2-\mu_1
	\cos\theta_1\big)\big(\mu_2+4\mu_3-\mu_1\cos\theta_1\big)\sin^2\xi\big)+\mu_3^2
	\sin^2(2\xi)
	\nonumber\\
	&-2\sqrt{2}\big(\mu_2+2\mu_3-\mu_1\cos\theta_1\big)\sin^2(2\xi)\bigg)^2\bigg(2
	\big(\mu_2-\mu_1\cos\theta_1\big)\big(\mu_2+4\mu_3-\mu_1\cos\theta_1\big)\cos^3\xi
	\sin\xi
	\nonumber\\
	&-2\cos\xi\sin\xi    
	\big(\mu_1^2\sin^2\theta_1+\big(\mu_2-\mu_1\cos\theta_1\big)\times\big(\mu_2
	+4\mu_3-\mu_1\cos\theta_1\big)\sin^2\xi\big)  + 2\mu_3^2\sin(4\xi) \bigg)   \, ,
\end{align*}
%%%%

%%%%
\begin{align*}
\mathcal{M}_{10}&=\bigg(2\cos^2\xi\big(\mu_1^2\sin^2\theta_1+\big(\mu_2+2\mu_3
-\mu_1\cos\theta_1\big)^2\sin^2\xi\big)\bigg)
\nonumber\\
&\bigg(2\big(\mu_2-\mu_1\cos\theta_1\big)\big(\mu_2+4\mu_3-\mu_1\cos\theta_1\big)
\cos^3\xi\sin\xi
\nonumber\\
&-2\cos\xi\sin\xi    \big(\mu_1^2\sin^2\theta_1+\big(\mu_2-\mu_1\cos\theta_1\big)
\times\big(\mu_2+4\mu_3-\mu_1\cos\theta_1\big)\sin^2\xi\big) 
\nonumber\\
& + 2\mu_3^2\sin(4\xi)-4\sqrt{2}\big(\mu_2+2\mu_3-\mu_1\cos\theta_1\big)\sin(4\xi)
\bigg)  \, ,
\end{align*}
%%%%

%%%%
\begin{align*}
	\mathcal{M}_{11}&=\bigg(64E+\cos^2\xi\big(\mu_1^2\sin^2\theta_1+\big(\mu_2
	-\mu_1\cos\theta_1\big)\big(\mu_2+4\mu_3-\mu_1\cos\theta_1\big)\sin^2\xi\big)
	+\mu_3^2\sin^2(2\xi)
	\nonumber\\
	&-2\sqrt{2}\big(\mu_2+2\mu_3-\mu_1\cos\theta_1\big)\sin^2(2\xi)\bigg)\bigg(2
	\big(\mu_2-\mu_1\cos\theta_1\big)\big(\mu_2+4\mu_3-\mu_1\cos\theta_1\big)\cos^3\xi
	\sin\xi
	\nonumber\\
	&-2\cos\xi\sin\xi    
	\big(\mu_1^2\sin^2\theta_1+\big(\mu_2-\mu_1\cos\theta_1\big)\times\big(\mu_2+4\mu_3
	-\mu_1\cos\theta_1\big)\sin^2\xi\big)  + 
	2\mu_3^2\sin(4\xi)\nonumber\\&-4\sqrt{2}\big(\mu_2+2\mu_3-\mu_1\cos\theta_1\big)
	\sin(4\xi) \bigg)  \, .
\end{align*}
%%%%

%%%%%%%%%%%%%%%%%%%%%%%%%%%%%%%%%%%%%%%%%%%

\end{document}